\documentclass[a4paper,12pt]{article}

\usepackage{amsmath, amsthm, amsfonts, amssymb}

\theoremstyle{plain}
\newtheorem{theorem}{Theorem}[section]

\newtheorem{corollary}[theorem]{Corollary}

\newtheorem{lemma}{Lemma}[section]

\theoremstyle{remark}
\newtheorem{remark}{Remark}[section]

\def\Om{\Omega}

\def\e{\varepsilon}

\def\G{\Gamma}
\def\l{\lambda}
\def\p{\partial}
\def\D{\Delta}

\def\k{\kappa}
\def\E{\mbox{\rm e}}
\def\a{\alpha}
\def\b{\beta}
\def\si{\sigma}

\def\d{\delta}

\def\vp{\varphi}

\def\vr{\varrho}

\def\Odr{\mathcal{O}}
\def\H{W_2}

\def\Ho{W_{2,0}}

\def\di{\,\mathrm{d}}

\def\I{\mathrm{I}}
\def\iu{\mathrm{i}}

 \DeclareMathOperator{\RE}{Re}
\DeclareMathOperator{\IM}{Im} \DeclareMathOperator{\spec}{\sigma}
\DeclareMathOperator{\discspec}{\sigma_{disc}}
\DeclareMathOperator{\essspec}{\sigma_{ess}}

\DeclareMathOperator{\dist}{dist}

\DeclareMathOperator{\dvr}{div}

\numberwithin{equation}{section}

\begin{document}
\allowdisplaybreaks

\title{Distant perturbations of the Laplacian in a multi-dimensional
space}

\author{Denis I. Borisov}
\date{}
\maketitle

\begin{abstract}
We consider the Laplacian in $\mathbb{R}^n$ perturbed by a
finite number of distant perturbations those are abstract
localized operators. We study the asymptotic behaviour of the
discrete spectrum as the distances between perturbations tend to
infinity. The main results are the convergence theorem and the
asymptotics expansions for the eigenelements. Some examples of
the possible distant perturbations are given; they are
potential, second order differential operator, magnetic
Schr\"odinger operator, integral operator, and $\d$-potential.
\end{abstract}

\begin{quote}
{\small { Nuclear Physics Institute, Academy of Sciences, 25068
\v Re\v z near Prague, Czechia
\\
Bashkir State Pedagogical University, October Revolution
St.~3a,
\\
 450000 Ufa, Russia
}
\\
E-mail: \texttt{borisovdi@yandex.ru}}
\end{quote}

\section*{Introduction}

Spectra of self-adjoint operators with distant perturbations
exhibit various interesting features and such operators were
studied quite intensively. Much attention was paid to a multiple
well Shr\"odinger operator in the case the wells were separated
by a large distance (see, for instance, \cite{H2, KS, KM, H1},
\cite[Sec. 8.6]{D}). The similar problem for the Dirac operator
was treated in \cite{HK}.  The main result of the cited works
was the description of the asymptotics behaviour of the isolated
eigenvalues as the distances between wells tend to infinity.
Recently new problems with more complicated distant
perturbations have been considered. S.~Kondej and I.~Veseli\'c
studied a $\d$-potential supported by a curve which consists of
a several components \cite{KV}. 
In the case these components are separated by a large distance
their results imply an asymptotic estimate for the lowest
spectral gap. 
The problems with distant perturbations were considered also for
the waveguides. In \cite{BE} the Dirichlet Laplacian in a planar
strip was studied, and the distant perturbations were two
segments of the same length on the boundary on which the
boundary condition switched to the Neumann one. The asymptotics
expansions for the isolated eigenvalues were constructed as the
distance between Neumann segments increased unboundedly. These
results were generalized in \cite{BE2}, where we studied
Dirichlet Laplacian in a domain formed by two adjacent strips of
arbitrary width coupled by two windows. These windows were
segments cut out from the common boundary of these strips. The
technique employed in \cite{BE2} followed the general ideas of
the paper \cite{B1}. In this paper we considered Dirichlet
Laplacian in an infinite multi-dimensional tube perturbed by two
distant perturbations. The perturbations were two abstract
localized operators. The asymptotics expansions for the
eigenvalues and the associated eigenfunctions were constructed.

In the present paper we consider the Laplacian in
$\mathbb{R}^n$, $n\geqslant 1$, perturbed by several distant
perturbations. The number of the perturbations is finite but
arbitrary and each perturbation is an abstract localized
operator. The restrictions for these operators are quite weak
and the results of this paper are applicable to a wide class of
distant perturbations of various nature (see Sec. 7).

In the paper we construct the asymptotics expansions for the
isolated eigenvalues and the associated eigenfunctions of the
problem considered. The technique we develop 
is a generalization of the approach employed in \cite{B1}. Such
generalization is needed since the tube considered in \cite{B1}
was infinite in one dimension only that is not the case for a
multi-dimensional space. The main additional ingredient we
involve is the technique borrowed from \cite[Ch. X\!I\!V, Sec.
4]{SP}. Our approach allows us actually to reduce the original
perturbed operator to a \emph{small regular} perturbation of the
direct sum of the limiting operators those are Laplacian with
one of the original perturbations. Due to this fact we believe
that this approach can be employed not only for the asymptotical
purposes, but also in studying other properties of the problems
with distant perturbations.

The structure of the paper is as follows. In the next section we
formulate the problem and present the main results. In the
second section we employ the technique from \cite[Ch. X\!I\!V,
Sec. 4]{SP} and transform the equation for the resolvent of the
both limiting and perturbed operators to a certain operator
equation. We employ it in the third section to obtain an
equation for the eigenelements of the perturbed operator. We
solve this equation explicitly using the slight modification of
the Birman-Schwinger approach suggested in \cite{G1}. This
allows us to prove the main results in the sixth section. The
seventh section is devoted to some examples of the distant
perturbation to which the general results of this paper can be
applied.


\section{Problem and main results}

Let $x=(x_1,\ldots,x_n)$ be the Cartesian coordinates in
$\mathbb{R}^n$, $n\geqslant 1$. Given any bounded domain
$Q\subset\mathbb{R}^n$  by $L_2(\mathbb{R}^n;Q)$ we denote the
subset of the functions from $L_2(\mathbb{R}^n)$ whose support
lies inside $\overline{Q}$.

Let $\Om_i\subset\mathbb{R}^n$, $i=1,\ldots,m$, be bounded
non-empty domains with infinitely differentiable boundary. By
$\mathcal{L}_i: \H^2(\Om_i)\to L_2(\mathbb{R}^n,\Om_i)$,
$i=1,\ldots,m$, we denote linear bounded operators satisfying
the relations
\begin{gather}
(\mathcal{L}_i u_1,u_2)_{L_2(\Om_i)}=(u_1,\mathcal{L}_i
u_2)_{L_2(\Om_i)}, \label{1.1}
\\
\big|(\mathcal{L}_i u,u)\big|\leqslant c_0\|\nabla
u\|_{L_2(\Om_i)}^2+c_1\|u\|_{L_2(\Om_i)}^2\label{1.2}
\end{gather}
for all $u, u_1,u_2\in \Ho^2(\Om_i)$, where $c_0$, $c_1$ are
some constants independent of $u$, $u_1$, $u_2$, and
\begin{equation}\label{1.3}
c_0<1.
\end{equation}
Since each $u\in\H^2(\mathbb{R}^n)$ belongs to
$\H^2(\mathbb{R}^n)$, we can regard $\H^2(\mathbb{R}^n)$ as a
subset of $\H^2(\Om_i)$. Due to such embedding we can define the
operators $\mathcal{L}_i$ on the space $\H^2(\mathbb{R}^n)$ and
consider the operators $\mathcal{L}_i$ as unbounded ones in
$L_2(\mathbb{R}^n)$.

%

We introduce the shift operator in $L_2(\mathbb{R}^n)$ as
$\mathcal{S}(a)u:=u(\cdot+a)$, where $a\in\mathbb{R}^n$. Let
$X_i$, $i=1,\ldots,m$, be some points in $\mathbb{R}^n$ and
denote $X:=(X_1,\ldots,X_m)$, $l_{i,j}:=|X_i-X_j|$. We set
\begin{equation*}
\mathcal{L}_X:=\sum\limits_{i=1}^{m}
\mathcal{S}(-X_i)\mathcal{L}_i \mathcal{S}(X_i).
\end{equation*}
This operator is defined on $\H^2(\Om_X)$,
$\Om_X:=\bigcup\limits_{i=1}^m\big(\Om_i+\{X_i\}\big)$,
$\Om_i+\{X_i\}:=\{x: x-X_i\in\Om_i\}$, and maps this space into
$L_2(\mathbb{R}^n;\Om_X)$. In what follows we assume that the
distances between $X_i$ increases unboundedly, i.e.,
$l_{i,j}\to+\infty$, $i\not=j$. Hence the distances between the
domains $\Om_i+\{X_i\}$ tend to infinity and the operator
$\mathcal{L}_X$ can be naturally treated as the distant
perturbation formed by the operators $\mathcal{L}_i$,
$i=1,\ldots,m$. We can also consider $\mathcal{L}_X$ as an
unbounded one in $L_2(\mathbb{R}^n)$ having $\H^2(\mathbb{R}^n)$
as the domain.

The main object of our study is the operator
$\mathcal{H}_X:=-\D_{\mathbb{R}^n}+\mathcal{L}_X$ in
$L_2(\mathbb{R}^n)$ with the domain $\H^2(\mathbb{R}^n)$. Here
$\D_{\mathbb{R}^n}$ denotes the Laplacian in $L_2(\mathbb{R}^n)$
with the domain $\H^2(\mathbb{R}^n)$. Our main aim is to study
the behaviour of the spectrum of $\mathcal{H}_X$ as
$l_{i,j}\to+\infty$.

Let $\mathcal{H}_i:=-\D_{\mathbb{R}^n}+\mathcal{L}_i$ be the
operators in $L_2(\mathbb{R}^n)$ having $\H^2(\mathbb{R}^n)$ as
the domain. Throughout the paper we assume that $\mathcal{H}_i$
and $\mathcal{H}_X$ are self-adjoint. By $\spec(\cdot)$,
$\essspec(\cdot)$, $\discspec(\cdot)$ we denote the spectrum,
the essential and the discrete spectrum of an operator.

Our first result is as follows.

\begin{theorem}\label{th1.1}
The essential spectra of $\mathcal{H}_i$, $\mathcal{H}_X$
coincide with the semi-axis $[0,+\infty)$. The discrete spectra
of these operator consist of finitely many negative eigenvalues.
The total multiplicity of the isolated eigenvalues of
$\mathcal{H}_X$ is bounded uniformly on $l_{i,j}$ provided these
lengths are large enough.
\end{theorem}

We denote
$\sigma_*:=\bigcup\limits_{i=1}^m\discspec(\mathcal{H}_i)$. We
say that $\l_*\in\sigma_*$ is $(p_1+\ldots+p_m)$-multiple if it
is a $p_i$-multiple eigenvalue of $\mathcal{H}_i$,
$i=1,\ldots,m$. The relation $p_i=0$ corresponds to the case
that $\l_*$ is not in the spectrum of $\mathcal{H}_i$.  Let
$l_X:=\min\limits_{i,j} l_{i,j}$.

\begin{theorem}\label{th1.2}
Each isolated eigenvalue of $\mathcal{H}_X$ converges to zero or
to $\l_*\in\sigma_*$ as $l_X\to+\infty$. If $\l_*\in\sigma_*$ is
$(p_1+\ldots p_m)$-multiple, the total multiplicity of the
eigenvalues of $\mathcal{H}_X$ converging to $\l_*$ equals
$p_1+\ldots p_m$.
\end{theorem}

\begin{theorem}\label{th1.3}
Let $\l_*\in\si_*$ be $(p_1+\ldots+p_m)$-multiple, and let
$\l_i=\l_i(X)\xrightarrow[l_X\to+\infty]{}\l_*$, $i=1,\ldots,p$,
$p:=p_1+\ldots+p_m$, be the eigenvalues of the operator
$\mathcal{H}_X$ taken counting multiplicity and ordered as
follows:
\begin{equation*}
0\leqslant |\l_1(X)-\l_*|\leqslant |\l_2(X)-\l_*|\leqslant\ldots
\leqslant |\l_p(X)-\l_*|.
\end{equation*}
These eigenvalues solve the equation (\ref{4.15}) and satisfy
the asymptotic formulas:
\begin{equation}\label{1.6}
\l_i(X)=\l_*+\tau_i(X)\left(1+\Odr\left(l_X^{-\frac{n-3}{2p}}
\E^{-l_X\frac{\sqrt{-\l_*}}{p}}\right)\right),\quad l\to+\infty.
\end{equation}
Here
\begin{equation*}
\tau_i=\tau_i(X)=\Odr\left(l_X^{-\frac{n-1}{2}}
\E^{-l_X\sqrt{-\l_*}}\right),\quad l_X\to+\infty,
\end{equation*}
are the zeroes of the polynomial $\det\big(\tau
\mathrm{E}-\mathrm{A}(\l_*,X)\big)$ taken counting multiplicity
and ordered as follows:
\begin{equation*}
0\leqslant |\tau_1(X)|\leqslant |\tau_2(X)|\leqslant\ldots
\leqslant |\tau_p(X)|.
\end{equation*}
The matrix $\mathrm{A}$ is defined by (\ref{4.12}). The
eigenfunctions associated with $\l_i$ obey the asymptotic
representation
\begin{gather*}
\psi_i=\sum\limits_{j=1}^{m} \mathcal{S}(-X_j)
\sum\limits_{q=1}^{p_j} \kappa_{\a_j+q}^{(i)} \psi_{j,q}
+\Odr\big(l_X^{-\frac{n-1}{2}}\E^{-l_X\sqrt{-\l_j}}\big), \quad
l_X\to+\infty,
\\
\a_1:=0,\quad \a_j:=p_1+\ldots+p_{j-1},
\end{gather*}
in $\H^2(\mathbb{R}^n)$-norm. Here $\psi_{i,j}$,
$j=1,\ldots,p_i$, are the eigenfunctions of $\mathcal{H}_i$
associated with $\l_*$ and are orthonormalized in
$L_2(\mathbb{R}^{n})$. The numbers $\kappa^{(i)}_j$ are the
components of the vectors
\begin{equation*}
\boldsymbol{\kappa}_i=\boldsymbol{\kappa}_i(X)=
\begin{pmatrix}
\kappa^{(i)}_1(X)
\\
\vdots
\\
\kappa^{(i)}_p(X)
\end{pmatrix},
\end{equation*}
which are the solutions to the system (\ref{4.11}) for
$\l=\l_i(X)$ and satisfy the condition
\begin{equation}\label{1.4}
(\boldsymbol{\kappa}_i,\boldsymbol{\kappa}_j)_{\mathbb{C}^p}=
\begin{cases}
1,&  i=j,
\\
\Odr\left(l_X^{\frac{n-1}{2}}\E^{-l_X\sqrt{-\l_*}}\right), &
i\not=j.\end{cases}
\end{equation}
\end{theorem}

As it is stated in this theorem, the leading terms of the
asymptotics expansions for the eigenvalues $\l_i$ are determined
by the matrix $\mathrm{A}(\l_*,X)$. At the same time it could be
a difficult problem to calculate  this matrix and its
eigenvalues explicitly. In the following theorems we show how to
calculate the asymptotics expansions for $\l_i$ in more explicit
form.

We will say that a square matrix $\mathrm{A}(X)$ satisfies the
condition (A) if it is diagonalizable and the determinant of the
matrix formed by the normalized eigenvectors of $\mathrm{A}(X)$
is separated from zero uniformly in $l_{i,j}$ large enough.

\begin{theorem}\label{th1.4}
Let the hypothesis of Theorem~\ref{th1.3} hold true and suppose
that the matrix $\mathrm{A}(\l_*,X)$ can be represented as
\begin{equation}\label{1.9}
\mathrm{A}(\l_*,X)=\mathrm{A}_0(X)+\mathrm{A}_1(X),
\end{equation}
where $\mathrm{A}_0$ satisfies the condition (A) and
$\|\mathrm{A}_1(X)\|\to0$ as $l_X\to+\infty$. Then the
eigenvalues $\l_i$ of $\mathcal{H}_X$ obey the asymptotic
formulas
\begin{equation*}
\l_i=\l_*+\tau_i^{(0)}\left(1+\Odr\left(l_X^{-\frac{n-3}{2}}
\E^{-l_X\sqrt{-\l_*}}\right)\right)+
\Odr(\|\mathrm{A}_1(l)\|),\quad l_X\to+\infty.
\end{equation*}
Here $\tau_i^{(0)}=\tau_i^{(0)}(X)$ are the roots of the
polynomial $\det\big(\tau\mathrm{E}-\mathrm{A}_0(X)\big)$ taken
counting multiplicity and ordered as follows:
\begin{equation}\label{1.11}
0\leqslant |\tau_1^{(0)}(X)|\leqslant
|\tau_2^{(0)}(X)|\leqslant\ldots \leqslant |\tau_p^{(0)}(X)|.
\end{equation}
Each of these roots satisfies the estimate
\begin{equation*}
\tau_i^{(0)}(X)=\Odr(\|\mathrm{A}_0(X)\|),\quad l_X\to+\infty.
\end{equation*}
\end{theorem}

We denote $X_{i,j}:=X_i-X_j$.

\begin{theorem}\label{th1.5}
Let the hypothesis of Theorem~\ref{th1.3} holds true. Then the
eigenvalues $\l_i$ satisfy the asymptotic formulas
\begin{equation}\label{1.12}
\l_i(X)=\l_*+\tau_i^{(0)}(X)+\Odr\left(l_X^{-n+2}
\E^{-2l_X\sqrt{-\l_*}}\right),\quad l_X\to+\infty.
\end{equation}
Here $\tau_i^{(0)}$ are the roots of the polynomial
$\det\big(\tau\mathrm{E}-\mathrm{A}_0\big)$ taken counting
multiplicity and ordered in accordance with (\ref{1.11}), and
the hermitian matrix $\mathrm{A}_0$ reads as follows:
\begin{equation*}
A_{i,j}^{(0)}(X):=\big(\mathcal{L}_k
\mathcal{S}(X_{k,r})\psi_{r,s},\psi_{k,q}\big)_{L_2(\Om_k)},
\quad \text{if}\quad k\not=r,\quad A_{i,j}^{(0)}(X):=0, \quad
\text{if}\quad k=r,
\end{equation*}
where $k=1,\ldots,m$, $q=1,\ldots,p_k$, $i=\a_k+q$,
$r=1,\ldots,m$, $s=1,\ldots,p_r$, $i=\a_r+s $. The estimates
\begin{equation*}
\tau_i^{(0)}=\Odr\left(l_X^{-\frac{n-1}{2}}
\E^{-l_X\sqrt{-\l_*}}\right),\quad l_X\to+\infty,
\end{equation*}
are valid.
\end{theorem}

\begin{corollary}\label{cr1.6}
Let $\l_*\in\sigma_*$ be $(1+1+\ldots+0)$-multiple, and
$\psi_i$, $i=1,2$ be the associated eigenfunctions of
$\mathcal{H}_i$ normalized in $L_2(\mathbb{R}^n)$. Then the
asymptotics expansions for the eigenvalues $\l_i$, $i=1,2$, are
as follows
\begin{align*}
&\l_1=\l_*-\left|\big(\mathcal{L}_1
\mathcal{S}(X_{1,2})\psi_2,\psi_1\big)_{L_2(\Om_1)}\right|
+\Odr\left(l_X^{-n+2} \E^{-2l_X\sqrt{-\l_*}}\right),\quad
l_X\to+\infty,
\\
&\l_2=\l_*+\left|\big(\mathcal{L}_1
\mathcal{S}(X_{1,2})\psi_2,\psi_1\big)_{L_2(\Om_1)}\right|
+\Odr\left(l_X^{-n+2} \E^{-2l_X\sqrt{-\l_*}}\right),\quad
l_X\to+\infty.
\end{align*}
\end{corollary}

\begin{theorem}\label{th1.7}
Let $\l_*\in\sigma_*$ be $(1+0+\ldots+0)$-multiple, and $\psi_1$
be the associated eigenfunction of $\mathcal{H}_1$ normalized in
$L_2(\mathbb{R}^n)$. Then the asymptotic expansion for the
eigenvalue $\l(X)\xrightarrow[l_X\to+\infty]{}\l_*$ of
$\mathcal{H}_X$ reads as follows
\begin{equation*}
\l(X)=\l_*-\sum\limits_{j=2}^{m}\big( \mathcal{L}_1
\mathcal{S}(X_{1,j})(\mathcal{H}_j-\l_*)^{-1}\mathcal{L}_j
\mathcal{S}(X_{j,1})\psi_1,\psi_1\big)_{L_2(\Om_1)}
+\Odr\big(l_X^{-\frac{3n-5}{2}}\E^{-3l_X\sqrt{-\l_*}}\big)
\end{equation*}
as $l_X\to+\infty$. The associated eigenfunction $\psi$ satisfy
the asymptotic representation
\begin{equation*}
\psi(x,X)=\psi_1(x-X_1)+
\Odr\big(l_X^{-\frac{n-1}{2}}\E^{-l_X\sqrt{-\l_*}}\big),\quad
l_X\to+\infty.
\end{equation*}
\end{theorem}

\begin{remark}\label{rm1.1}
In this theorem the operators $(\mathcal{H}_j-\l_*)$,
$j=2,\ldots,m$, are boundedly invertible since
$\l_*\not\in\discspec(\mathcal{H}_j)$.
\end{remark}

In accordance with Theorem~\ref{th1.4} the leading terms of the
asymptotics expansions of the eigenvalues of $\mathcal{H}_X$ can
be expressed in terms of the matrix $\mathrm{A}_0$ once it is
possible to approximate $\mathrm{A}(\l_*,X)$ in the sense of
(\ref{1.9}). One of the possible ways to employ
Theorem~\ref{th1.4} is given in Theorem~\ref{th1.5}. Here the
matrix $\mathrm{A}_0$ is calculated explicitly in terms of the
limiting eigenfunctions and the operators $\mathcal{L}_i$. We
also observe that this matrix is in fact the first-order term in
the asymptotic expansion for $\mathrm{A}(\l_*,X)$.

One of the general cases is that the number $\l_*\in\sigma_*$ is
a simple isolated eigenvalue of two of operators
$\mathcal{H}_i$. This case is addressed in
Corollary~\ref{cr1.6}. We stress that in this case the
asymptotics expansions for the eigenvalues are very similar to
ones for a double-well Schr\"odinger operator with symmetric
wells (see, for instance, \cite[Th. 2.8]{H1}). At the same time,
in our case the number of distant perturbations is arbitrary and
no symmetry is assumed.

One more general case is that $\l_*$ is a simple isolated
eigenvalue of one of the operators $\mathcal{H}_i$ only. The
results for this case are due to Theorem~\ref{th1.7}. In this
case Theorem~\ref{th1.5} does not provide good asymptotics
expansions for the eigenvalues of $\mathcal{H}_X$ since the
matrix $\mathrm{A}_0$ in this theorem is zero. In view of this
fact we have to use second-order term of the asymptotic
expansion for $\mathrm{A}(\l_*,X)$. We also note that in this
case the leading terms in the asymptotics expansion for the
eigenvalues of $\mathcal{H}_X$ are smaller by order than leading
terms in (\ref{1.12}).

Generally speaking, some of the eigenvalues of the matrix
$\mathrm{A}_0$ in Theorem~\ref{th1.5} can be identically zero
for large $l_{i,j}$. In this case the leading terms in
(\ref{1.12}) vanish. If it occurs, one should employ
next-to-leading terms of the asymptotics expansion for
$\mathrm{A}(\l_*,X)$ and to treat them as a part of
$\mathrm{A}_0$ in (\ref{1.9}). Such an expansion for
$\mathrm{A}(\l_*,X)$ can be obtained by the technique employed
in the proof of Theorem~\ref{th1.7}. We do not provide such
results in the paper in order not to overload the text by quite
technical and bulky calculations.


\section{Proof of Theorem~\ref{th1.1}}

Let $\Om\subset \mathbb{R}^n$ be a bounded non-empty domain and
$\mathcal{L}: \H^2(\Om)\to L_2(\mathbb{R}^n;\Om)$ be an operator
satisfying the relations
\begin{equation}\label{2.1}
(\mathcal{L} u_1,u_2)_{L_2(\Om)}=(u_1,\mathcal{L}
u_2)_{L_2(\Om)}, \quad \big|(\mathcal{L} u,u)\big|\leqslant
c_0\|\nabla u\|_{L_2(\Om_\pm)}^2+c_1\|u\|_{L_2(\Om)}^2
\end{equation}
for all $u,u_1,u_2\in\H^2(\Om)$, where $c_0$, $c_1$ are some
constants, and $c_0$ obeys (\ref{1.3}). We introduce the
operator
$\mathcal{H}_{\mathcal{L}}:=-\D_{\mathbb{R}^n}+\mathcal{L}$ in
$L_2(\mathbb{R})$ with the domain $\H^2(\mathbb{R})$ and assume
that it is self-adjoint.

\begin{lemma}\label{lm2.1}
$\essspec(\mathcal{H}_{\mathcal{L}})=[0,+\infty)$.
\end{lemma}

\begin{proof}
We will employ Weyl criterion to prove the lemma. Let
$\l\in[0,+\infty)$. By $\chi=\chi(t)$ we denote an infinitely
differentiable function cut-off function being one as $r<0$ and
vanishing as $r>1$. We introduce the sequence of the functions
$u_p(x):=c_p |x|^{-n/2+1} J_{n/2-1}(\sqrt{\l}|x|)\chi(|x|-p)\in
\H^2(\mathbb{R}^n)$, where $J_q$ is the Bessel function of
$q$-th order. The coefficients $c_p$ are specified by the
normalization condition $\|u_p\|_{L_2(\mathbb{R}^n)}=1$. Since
\begin{equation*}
|x|^{-n+2}J_{n/2-1}^2(\sqrt{\l}|x|)=\frac{2|x|^{-n+1}}{\pi
\sqrt{\l}} \left(\cos^2\left(\sqrt{\l}|x|-
\frac{(n-3)\pi}{4}\right)+\Odr(|x|^{-1})\right),
\end{equation*}
as $|x|\to+\infty$, it follows that
$c_p\xrightarrow[p\to+\infty]{}0$. Using this fact it is easy to
check that $\|\mathcal{L} u_p\|_{L_2(\mathbb{R}^n)}\to0$,
$\|\mathcal{H}_{\mathcal{L}} u_p\|_{L_2(\mathbb{R}^n)}\to0$ as
$p\to+\infty$. Therefore, $u_p$ is a singular sequence for
$\mathcal{H}_{\mathcal{L}}$ at $\l$ and
$[0,+\infty)\subseteq\essspec(\mathcal{H}_{\mathcal{L}})$. The
opposite inclusion can be shown completely by analogy with how
the same was established in the proof of Lemma~2.1 in \cite{B1}.
\end{proof}

\begin{lemma}\label{lm2.2}
The discrete spectrum of the operator
$\mathcal{H}_{\mathcal{L}}$ consists of finitely many negative
eigenvalues.
\end{lemma}

The proof of this lemma is the same as that of Lemma~2.2 in
\cite{B1}.

We apply now Lemmas~\ref{lm2.1},~\ref{lm2.2} with
$\mathcal{L}=\mathcal{L}_i$, $\Om=\Om_i$, $i=1,\ldots,m$ and
arrive at the statement of the theorem on $\mathcal{H}_i$. It
also follows from Lemmas~\ref{lm2.1},~\ref{lm2.2} with
$\mathcal{L}=\mathcal{L}_X$, $\Om:=\Om_X$, that the essential
spectrum of $\mathcal{H}_X$ coincides with $[0,+\infty)$ and the
discrete spectrum consists of finitely many eigenvalues. It
remains to check that the total multiplicity of these
eigenvalues is independent on $l_{i,j}$ provided these lengths
are large enough. Completely in the same way how the estimate
(2.5) was established in the proof of Lemma~2.2 in \cite{B1},
one can check that
\begin{equation}\label{2.2}
\mathcal{H}_X\geqslant \mathcal{H}_X^{(0)}\oplus
\mathcal{H}_X^{(1)},
\end{equation}
where $\mathcal{H}_X^{(1)}$ is the negative Neumann Laplacian in
$\mathbb{R}^n\setminus\Om_X$, while $\mathcal{H}_X^{(0)}$
denotes the operator
\begin{equation*}
-\dvr\left(1-c_0\sum\limits_{i=1}^{m}
\chi(|x-X_i|-\e)\right)\nabla-c_1\sum\limits_{i=1}^{m}
\chi(|x-X_i|-\e)
\end{equation*}
in $\Om_X$ subject to Neumann boundary condition.  Here $\e$ is
such that $\Om_i\subseteq\{x: |x|<\e\}$, and the lengths
$l_{i,j}$ are supposed to be large enough so that supports of
$\chi(|x-X_i|-\e)$ do not intersect for different $i$. It is
clear that $\mathcal{H}_X^{(0)}$ is unitary equivalent to the
sum $\bigoplus\limits_{i=1}^m \mathcal{H}_{X_i}^{(0)}$, where
$\mathcal{H}_{X_i}^{(0)}$ is the operator
\begin{equation*}
-\dvr\big(1-c_0 \chi(|x-X_i|-\e)\big)\nabla-c_1 \chi(|x-X_i|-\e)
\end{equation*}
in $\{x: |x|<\e\}$ subject to Neumann boundary condition. This
sum is independent on $l_{i,j}$ and has a finite number of
negative isolated eigenvalues. By the minimax principle and
(\ref{2.2}) these eigenvalues give the lower bounds for the
negative eigenvalues of $\mathcal{H}_X$ that implies that total
multiplicity of the negative eigenvalues of $\mathcal{H}_X$ is
bounded uniformly on $l_{i,j}$ provided these quantities are
large enough.


\section{Reduction to an operator equation}

In this section we collect some preliminaries which will be
employed in the proof of Theorems~\ref{th1.2}-\ref{th1.7}.

Let $\mathcal{L}$ and $\mathcal{H}_{\mathcal{L}}$ be the
operators introduced in the previous section. For any $\e>0$ by
$\mathbb{S}_\e$ we indicate the set of complex numbers separated
from the half-line $[0,+\infty)$ by a distance greater than
$\e$. We also assume that $\e$ is chosen so that
$\discspec(\mathcal{H})\subset\mathbb{S}_\e$.

Consider the equation
\begin{equation}\label{3.1}
(\mathcal{H}_{\mathcal{L}}-\l)u=f,
\end{equation}
where $f\in L_2(\mathbb{R}^n;\Om^\b)$, $\Om^\b:=\{x\in
\mathbb{R}^n: \dist(\Om,x)<\b\}$, $\b>0$, $\l\in\mathbb{S}_\e$.
We are going to reduce this equation to an operator equation in
$L_2(\mathbb{R}^n;\Om^\b)$. In order to do it, we will employ
the general scheme borrowed from \cite[Ch. X\!I\!V, Sec. 4]{SP}.

Let $g\in L_2(\mathbb{R}^n;\Om^\b)$ be an arbitrary function. We
introduce $v:=(-\D_{\mathbb{R}^n}-\l)^{-1}g$. The function $v$
can be represented as
\begin{gather}
v(x,\l):=\int\limits_{\Om^\b} G_n(|x-y|,\l) g(y)\di
y,\label{3.2}
\\
G_n(t,\l):=-\frac{\iu^{\frac{n}{2}}
(\sqrt[4]{-\l})^{n-2}}{2^{\frac{n}{2}+1}\pi^{\frac{n}{2}-1}}
t^{-\frac{n}{2}+1}H_{\frac{n}{2}-1}^{(1)}(\iu
t\sqrt{-\l}),\nonumber
\end{gather}
where $H_{n/2-1}^{(1)}$ is the Hankel function of the first kind
and $(n/2-1)$-th order. The branches of the roots are specified
by the requirements $\RE\sqrt{-\l}>0$, $\RE\sqrt[4]{-\l}>0$,
$\IM\sqrt[4]{-\l}>0$ as $\l\in \mathbb{S}_\e$. 

We denote by $\mathcal{H}_\Om$ the operator $-\D+\mathcal{L}$ in
$L_2(\Om^\b)$ with domain $\Ho^2(\Om^\b)$. Here $\Ho^2(\Om^\b)$
consists of the functions from $\H^2(\Om^\b)$ vanishing on
$\p\Om^\b$. The operator $\mathcal{H}_\Om$ is symmetric (see
(\ref{2.1})), and the operator $(\mathcal{H}_\Om-\iu)^{-1}$ is
therefore well-defined and is bounded as an operator in
$L_2(\Om^\b)$. Moreover, $\mathcal{H}_\Om$ is bounded as an
operator from $\Ho^2(\Om^\b)$ into $L_2(\Om^\b)$. By Banach
theorem on inverse operator two last facts imply that the
operator $(\mathcal{H}_\Om-\iu)^{-1}: L_2(\Om^\b)\to
\Ho^2(\Om^\b)$ is bounded. Using this operator, we define one
more function $w:=-(\mathcal{H}_\Om-\iu)^{-1}\mathcal{L}v$.

By $\chi_\Om=\chi_\Om(x)$ we indicate infinitely differentiable
cut-off function being one in $\Om^{\b/2}$ and vanishing outside
$\Om^\b$. We construct the solution to the equation (\ref{3.1})
as
\begin{equation}\label{3.3}
u(x,\l)=\mathcal{T}_1(\l)g:=v(x,\l)+\chi_\Om(x) w(x,\l).
\end{equation}
This function is obviously an element of $\H^2(\mathbb{R}^n)$.
Now we apply the operator $(\mathcal{H}_{\mathcal{L}}-\l)$ to
this function:
\begin{align}
&(\mathcal{H}_{\mathcal{L}}-\l)u=g+\mathcal{L}v+
(-\D-\l+\mathcal{L})\chi_\Om w=g+\mathcal{T}_2(\l)g,\label{3.8}
\\
&\mathcal{T}_2(\l)g:=-2\nabla\chi_\Om\cdot\nabla
w-w(\D+\l-\iu)\chi_\Om.\nonumber
\end{align}
Here we have also used the identities $\mathcal{L}\chi_\Om
w=\mathcal{L} w=\chi_\Om \mathcal{L}w$. Thus, the equation
(\ref{3.1}) holds true if
\begin{equation}\label{3.4}
g+\mathcal{T}_2(\l)g=f.
\end{equation}

\begin{lemma}\label{lm3.1}
The operator $\mathcal{T}_1(\l):
L_2(\Om^\b)\to\H^2(\mathbb{R}^n)$ is bounded and holomorphic
w.r.t. $\l\in \mathbb{S}_\e$. The operator $\mathcal{T}_2(\l)$
is bounded in $L_2(\Om^\b)$ and holomorphic w.r.t.
$\l\in\mathbb{S}_\e$. For each solution of (\ref{3.4}) the
function $u$ defined by (\ref{3.3}) solves (\ref{3.1}). And vice
versa, for each solution $u$ of (\ref{3.1}) there exists unique
solution $g$ of (\ref{3.4}) satisfying the relation
$u=\mathcal{T}_1(\l)g$. This equivalence holds true for all
$\l\in\mathbb{S}_\e$.
\end{lemma}

\begin{proof}
The operator $(-\D_{\mathbb{R}^n}-\l)^{-1}:
L_2(\mathbb{R}^n;\Om^\b)\to\H^2(\mathbb{R}^n)$ is bounded and
holomorphic w.r.t. $\l\in\mathbb{S}_\e$ that can be established
by analogy with the proof of Lemma~3.1 in \cite{B1}. Since
$(\mathcal{H}_\Om-\iu)^{-1}\mathcal{L}$ is a bounded operator in
$\Ho^2(\Om^\b)$, we conclude that the mapping $g\mapsto w$ is a
bounded operator from $L_2(\mathbb{R}^n;\Om^\b)$ into
$\Ho^2(\Om^\b)$ being holomorphic w.r.t. $\l\in \mathbb{S}_\e$.
Thus, the operator $\mathcal{T}_1(\l):
L_2(\Om^\b)\to\H^2(\mathbb{R}^n)$ is bounded and holomorphic
w.r.t. $\l\in\mathbb{S}_\e$. This fact and the definition of
$\mathcal{T}_2$ imply that this operator is bounded and
holomorphic w.r.t. $\l\in \mathbb{S}_\e$ as an operator 
in $L_2(\Om^\b)$. 

Let $g$ solve the equation (\ref{3.4}); as it was shown above in
this case the function $u$ defined by (\ref{3.3}) is a solution
to the equation (\ref{3.1}). Suppose now that $u$ solves
(\ref{3.1}). By direct calculations one can check that the
corresponding $v$, $w$ and $g$ are given by the formulas
\begin{equation}\label{3.5}
w:=(\D_{\Om^\b}^{D}+\iu)^{-1}\mathcal{L}u, \quad v:=u-\chi_\Om
w,\quad g=\mathcal{T}_1^{-1}(\l)u:=(-\D-\l)v,
\end{equation}
where $\D_{\Om^\b}^{D}$ is the
Dirichlet Laplacian in $\Om^\b$.
\end{proof}

\begin{lemma}\label{lm3.2}
The operator $(\I+\mathcal{T}_2)^{-1}$ is bounded and
meromorphic on $\l\in \mathbb{S}_\e$. The poles of this operator
are simple and coincide with the isolated eigenvalues of
$\mathcal{H}_{\mathcal{L}}$. For $\l$ close to a $p$-multiple
eigenvalue $\l_*$ of $\mathcal{H}_{\mathcal{L}}$ the
representation
\begin{equation}\label{3.6}
(\I+\mathcal{T}_2(\l))^{-1}=-\sum\limits_{i=1}^{p}
\frac{\phi_i(\cdot,\psi_i)_{L_2(\Om^\b)}}{\l-\l_*}+
\mathcal{T}_3(\l)
\end{equation}
holds true. Here $\psi_i$ are the eigenfunctions associated with
$\l_*$ and orthonormalized in $L_2(\mathbb{R}^n)$,
$\phi_i:=\mathcal{T}_1^{-1}(\l_*)\psi_i$, and the operator
$\mathcal{T}_3: L_2(\Om^\b)\to L_2(\Om^\b)$ is bounded and
holomorphic w.r.t. $\l$ close to $\l_*$ as an operator in
$L_2(\Om^\b)$. The equation (\ref{3.4}) with $\l=\l_*$ is
solvable if and only if
\begin{equation}\label{3.7}
(f,\psi_i)_{L_2(\Om^\b)}=0,\quad i=1,\ldots,p,
\end{equation}
and the solution reads as follows
\begin{equation}\label{3.12}
g=\mathcal{T}_3(\l_*)f+\sum\limits_{j=1}^{m} c_i\phi_i,
\end{equation}
where $c_i$ are arbitrary constants.
\end{lemma}

\begin{proof}
It follows from (\ref{3.3}), (\ref{3.8}) that
$(\mathcal{H}_{\mathcal{L}}-\l)\mathcal{T}_1(\l)=
\I+\mathcal{T}_2(\l)$. Therefore,
\begin{equation}\label{3.9}
(\mathcal{H}_{\mathcal{L}}-\l)^{-1}=\mathcal{T}_1(\l)
(\I+\mathcal{T}_2(\l))^{-1}, \quad
(\I+\mathcal{T}_2(\l))^{-1}=\mathcal{T}_1^{-1}(\l)
(\mathcal{H}_{\mathcal{L}}-\l)^{-1},
\end{equation}
where the operator $\mathcal{T}_1^{-1}(\l)$ is defined by the
formulas (\ref{3.5}). By analogy with the proof of Lemma~3.1 in
\cite{B1} one can show that the operator
$(\mathcal{H}_{\mathcal{L}}-\l)^{-1}:
L_2(\mathbb{R}^n;\Om^\b)\to\H^2(\mathbb{R}^n)$ is meromorphic on
$\l\in\mathbb{S}_\e$, its poles coincide with the isolated
eigenvalues of $\mathcal{H}_{\mathcal{L}}$, and for $\l$ close
to $\l_*$ the representation
\begin{equation}\label{3.10}
(\mathcal{H}_{\mathcal{L}}-\l)^{-1}=-\sum\limits_{i=1}^{p}
\frac{\psi_i(\cdot,\psi_i)_{L_2(\mathbb{R}^n)}}{\l-\l_*}
+\mathcal{T}_4(\l)
\end{equation}
holds true, where the operator $\mathcal{T}_4(\l):
L_2(\mathbb{R}^n)\to \H^2(\mathbb{R}^n)$ is bounded and
holomorphic w.r.t. $\l$ close to $\l_*$. Hence, in view of
(\ref{3.9}), (\ref{3.10}), and (\ref{3.5}), the operator
$(\I+\mathcal{T}_2)^{-1}$ is meromorphic on
$\l\in\mathbb{S}_\e$, the poles of this operator are simple and
coincide with the isolated eigenvalues of ${H}_{\mathcal{L}}$,
and the representation (\ref{3.6}) holds true. As it also
follows from (\ref{3.10}), the equation (\ref{3.1}) with
$\l=\l_*$ is solvable if and only if  the relations (\ref{3.7})
are valid, and the solution of (\ref{3.1}) with $\l=\l_*$ is
given by the formula
$u=\mathcal{T}_4(\l_*)f+\sum_{j=1}^{p}c_i\psi_i$, where $c_i$
are arbitrary constants. Employing now Lemma~\ref{lm3.1}, we
conclude that the relations (\ref{3.7}) are the solvability
conditions for the equation (\ref{3.5}) with $\l=\l_*$. Thus,
the solution of this equation is defined uniquely up to a linear
combination of the functions $\phi_i$, $i=1,\ldots,m$. The
formula (\ref{3.12}) is valid since for the functions $f$
satisfying (\ref{3.7}) the identity
$\big(\I+\mathcal{T}_2(\l_*)\big) \mathcal{T}_3(\l_*)f=f$ holds
true  due to (\ref{3.6}).
\end{proof}

Let $\widetilde{\Om}\subset\mathbb{R}^n$ be a bounded domain
with infinitely differentiable boundary, and
$\widetilde{X}\in\mathbb{R}^n$ be a point. Suppose that
$l:=|\widetilde{X}|$ is a large parameter. We define the
operator $\mathcal{T}_5(\l,\widetilde{X}):
L_2(\mathbb{R}^n;\Om^\b)\to\H^2(\widetilde{\Om})$ as follows
\begin{equation*}
\mathcal{T}_5(\l,\widetilde{X}):=
\mathcal{S}(\widetilde{X})(-\D_{\mathbb{R}^n}-\l)^{-1}.
\end{equation*}

\begin{lemma}\label{lm3.4}
The operator $\mathcal{T}_5$ is bounded and holomorphic on
$\l\in\mathbb{S}_\e$. For any compact set
$\mathbb{K}\in\mathbb{S}_\e$ the estimates
\begin{equation}\label{3.11}
\left\|\frac{\p^i \mathcal{T}_5}{\p\l^i}\right\|\leqslant C
l^{-\frac{n-2i-1}{2}}\E^{-l\sqrt{-\l}},\quad i=0,1,
\end{equation}
hold true, where the constant $C$ is independent on
$\widetilde{X}$ and $\l\in\mathbb{K}$.
\end{lemma}

\begin{proof}
As it was said in the proof of Lemma~\ref{lm3.2} the operator
$(-\D_{\mathbb{R}^n}-\l)^{-1}:
L_2(\mathbb{R}^n;\Om^\b)\to\H^2(\mathbb{R}^n)$ is bounded and
holomorphic w.r.t. $\l\in\mathbb{S}_\e$. Therefore, the same is
true for the operator $\mathcal{T}_5$. The estimates
(\ref{3.11}) follow from the asymptotics 
\begin{equation}\label{3.16}
G_n(t,\l)=-\frac{(\sqrt[4]{-\l})^{n-3}}{2^{(n+1)/2}
\pi^{(n-1)/2}} t^{-(n-1)/2}\E^{-t\sqrt{-\l}}
\Big(1+\Odr\big(|\l|^{-1/2}t^{-1}\big)\Big),
\end{equation}
as $t\to+\infty$, $\l\in\mathbb{S}_\e$; this formula can be
differentiated w.r.t. $\l$.
\end{proof}


\section{Equation for the eigenelements of $\mathcal{H}_X$}

In this section we will obtain the equation for the eigenvalues
and the eigenfunctions of the operator $\mathcal{H}_X$ and will
solve this equation explicitly.

By $\mathcal{T}_j^{(i)}$, $\mathcal{T}_j^{(X)}$, we denote the
operators $\mathcal{T}_j$ from the previous section
corresponding to $\mathcal{L}=\mathcal{L}_j$, $\Om=\Om_j$,
$\mathcal{L}=\mathcal{L}_X$, $\Om=\Om_X$. Let us study the
structure of the operator $\mathcal{T}_2^{(X)}$ in more details.

Given $g\in L_2(\Om_X^\b)$, due to (\ref{3.2}) we have
\begin{gather}
\begin{aligned}
v_X(x,\l)&=\int\limits_{\Om_X^\b}G_n(|x-t|,\l)g(t)\di t=
\sum\limits_{i=1}^{m}\int\limits_{\Om_i^\b+\{X_i\}}
G_n(|x-t|,\l)g(t)\di t
\\
&=\sum\limits_{i=1}^{m}\int\limits_{\Om_i^\b}
G_n(|x-X_i-t|,\l)g_i(t)\di t=\sum\limits_{i=1}^{m}
\big(\mathcal{S}(-X_i)v_i\big)(x,\l),
\end{aligned}\label{4.1}
\\
g_i(t):=g(X_i+t),\quad v_i(x):=\int\limits_{\Om_i^\b}
G_n(|x-t|,\l)g_i(t)\di t.\nonumber
\end{gather}
Now we apply the operator $\mathcal{L}_X$ to the function $v_X$
and obtain:
\begin{gather}
\mathcal{L}_X v_X=\sum\limits_{i=1}^{m} \mathcal{S}(-X_i)
\mathcal{L}_i\left(
v_i+\sum\limits_{\genfrac{}{}{0pt}{}{j=1}{j\not=i}}^{m}
\mathcal{S}(X_{i,j})v_j \right)=\sum\limits_{i=1}^{m}
\mathcal{S}(-X_i)
\mathcal{L}_i\left(v_i+\widetilde{v}_i\right),\label{4.2}
\\
\widetilde{v}_i:=
\sum\limits_{\genfrac{}{}{0pt}{}{j=1}{j\not=i}}^{m}
\mathcal{S}(X_{i,j})v_j =
\sum\limits_{\genfrac{}{}{0pt}{}{j=1}{j\not=i}}^{m}
\mathcal{T}_5(\l,X_{i,j})g_j.\nonumber
\end{gather}
We introduce the functions
\begin{align*}
&w_i:=-(\mathcal{H}_{\Om_i}-\iu)^{-1} \mathcal{L}_i v_i, \quad
\widetilde{w}_i:=-(\mathcal{H}_{\Om_i}-\iu)^{-1} \mathcal{L}_i
\widetilde{v}_i,
\\
&w_X:=W_X+\widetilde{W}_X,\quad W_X:=\sum\limits_{i=1}^{m}
\mathcal{S}(-X_i)w_i,\quad
\widetilde{W}_X:=\sum\limits_{i=1}^{m} \mathcal{S}(-X_i)
\widetilde{w}_i.
\end{align*}
It is obvious that $w_X, W_X, \widetilde{W}_X\in
\Ho^2(\Om_X^\b)$. Since $\mathcal{L}_X w_X=\sum\limits_{i=1}^{m}
\mathcal{S}(-X_i) \mathcal{L}_i (w_i+\widetilde{w}_i)$, we
obtain
\begin{align*}
(\mathcal{H}_{\Om_X}-\iu) w_X&=\sum\limits_{i=1}^{m}
\Big(-(\Delta+\iu)\mathcal{S}(-X_i)
+\mathcal{S}(-X_i)\mathcal{L}_i\Big)(w_i+\widetilde{w}_i)
\\
&
=-\sum\limits_{i=1}^{m}
\mathcal{S}(-X_i)\mathcal{L}_i
(v_i+\widetilde{v}_i)=-\mathcal{L}_X v_X,
\\
w_X&=-
(\mathcal{H}_{\Om_X}-\iu)^{-1}
\mathcal{L}_X v_X.
\end{align*}
We define the cut-off function $\chi_{\Om_X}:=\sum_{i=1}^m
\mathcal{S}(-X_i)\chi_{\Om_i}$, where the function
$\chi_{\Om_i}$ corresponds to the operator
$\mathcal{T}_1^{(i)}$. In this case the operator
$\mathcal{T}_1^{(X)}$ reads as follows:
\begin{align*}
\mathcal{T}_1^{(X)}g&=\sum\limits_{i=1}^{m}
\mathcal{S}(-X_i)v_i+ \sum\limits_{i=1}^{m}
\mathcal{S}(-X_i)\chi_{\Om_i}(w_i+\widetilde{w}_i)
\\
&= \sum\limits_{i=1}^{m} \mathcal{S}(-X_i)(v_i+ \chi_{\Om_i}
w_i+\chi_{\Om_i}\widetilde{w}_i)=\sum\limits_{i=1}^{m}
\mathcal{S}(-X_i)
(\mathcal{T}_1^{(i)}g_i+\chi_{\Om_i}\widetilde{w}_i).
\end{align*}
Therefore,
\begin{align}
&\mathcal{T}_2^{(X)}(\l,X)g=\sum\limits_{i=1}^{m}
\mathcal{S}(-X_i) \mathcal{T}_2^{(i)}(\l)g_i +
\sum\limits_{i=1}^{m} \mathcal{S}(-X_i)
\sum\limits_{\genfrac{}{}{0pt}{}{j=1}{j\not=i}}^{m}
\mathcal{T}_6^{(i,j)}(\l) g_j,\label{4.5}
\\
&\mathcal{T}_6^{(i,j)}(\l):=\left(
2\nabla\chi_{\Om_i}\cdot\nabla
+(\D\chi_{\Om_i}+(\l-\iu)\chi_{\Om_i})\right)
(\mathcal{H}_{\Om_i}-\iu)^{-1}\mathcal{L}_i
\mathcal{T}_5(\l,X_{i,j}).\nonumber
\end{align}

\begin{lemma}\label{lm4.1}
The operators $\mathcal{T}_6^{(i,j)}:
L_2(\mathbb{R}^n;\Om_j^\b)\to L_2(\Om_i^\b)$ are bounded and
holomorphic w.r.t. $\l\in \mathbb{S}_\e$. The relation
\begin{equation}\label{4.5a}
\mathcal{T}_6^{(i,j)}(\l)=\mathcal{L}_i
\mathcal{T}_5(\l,X_{i,j})+
(\D-\mathcal{L}_i+\l)\chi_{\Om_i}(\mathcal{H}_{\Om_i}-\iu)^{-1}
\mathcal{L}_i \mathcal{T}_5(\l,X_{i,j})
\end{equation}
is valid. For each compact set $\mathbb{K}\subset \mathbb{S}_\e$
the estimates
\begin{equation*}
\bigg\|\frac{\p^k \mathcal{T}_6^{(i,j)}}{\p\l^k}\bigg\|\leqslant
C l_{i,j}^{-\frac{n-2k-1}{2}}\E^{-l_{i,j}\sqrt{-\l}},\quad
k=0,1,
\end{equation*}
hold true, where the constant $C$ is independent on $l_{i,j}$
and $\l\in\mathbb{K}$.
\end{lemma}

The statement of the lemma follows from the definition of
$\mathcal{T}_6^{(i,j)}$ and Lemma~\ref{lm3.4}.

According to Lemma~\ref{lm3.1}, the eigenvalues of the operator
$\mathcal{H}_X$ are numbers for which the equation (\ref{3.4})
with $\mathcal{T}_2=\mathcal{T}_2^{(X)}$ and $f=0$ has a
nontrivial solution. Let $g_X$ be a solution to this equation.
Since $g_X=\sum\limits_{i=1}^{m} \mathcal{S}(-X_i)g_i$, due to
(\ref{4.5}) we conclude that the equation (\ref{3.4}) for $g_X$
can be rewritten as
\begin{equation*}
\sum\limits_{i=1}^{m} \mathcal{S}(-X_i)
\left(g_i+\mathcal{T}_2^{(i)}(\l)g_i+
\sum\limits_{\genfrac{}{}{0pt}{}{j=1}{j\not=i}}^{m}
\mathcal{T}_6^{(i,j)}(\l) g_j\right)=0.
\end{equation*}
Each term in this equation has a compact support and these
supports do not intersect if $l_{i,j}$ are large enough. Thus,
the obtained equation is equivalent to
\begin{equation}\label{4.6}
g_i+\mathcal{T}_2^{(i)}(\l)g_i+
\sum\limits_{\genfrac{}{}{0pt}{}{j=1}{j\not=i}}^{m}
\mathcal{T}_6^{(i,j)}(\l) g_j=0,\quad i,j=1,\ldots,m.
\end{equation}
We introduce two operators in the space
$\boldsymbol{L}:=\bigoplus\limits_{i=1}^{m}
L_2(\mathbb{R}^n;\Om_i^\b)$,
\begin{gather*}
\mathcal{T}_7(\l)\boldsymbol{g}:=
\big(\mathcal{T}_2^{(1)}(\l)g_1,\ldots,
\mathcal{T}_2^{(m)}(\l)g_m\big),
\\
\mathcal{T}_8(\l,X)\boldsymbol{g}:= \left(
\sum\limits_{\genfrac{}{}{0pt}{}{j=1}{j\not=1}}^{m}
\mathcal{T}_6^{(1,j)}(\l) g_j,\ldots,
\sum\limits_{\genfrac{}{}{0pt}{}{j=1}{j\not=m}}^{m}
\mathcal{T}_6^{(m,j)}(\l) g_j\right),
\end{gather*}
where $\boldsymbol{g}:=(g_1,\ldots,g_m)\in\boldsymbol{L}$.
Employing these operators, we can rewrite the equation
(\ref{4.6}) as follows:
\begin{equation}\label{4.7}
\boldsymbol{g}+\mathcal{T}_7(\l)\boldsymbol{g}+
\mathcal{T}_8(\l,X)\boldsymbol{g}=0.
\end{equation}

Let $\l_*\in\sigma_*$ be $(p_1+\ldots+p_m)$-multiple, and
$\psi_{i,j}$, $i=1,\ldots,m$, $j=1,\ldots,p_i$, be the
associated eigenfunctions of $\mathcal{H}_i$ orthonormalized in
$L_2(\mathbb{R}^n)$. We denote $p:=p_1+\ldots+p_m$,
\begin{align*}
&\boldsymbol{\phi}_{\a_1+j}:=(\phi_{1,j},0,0,\ldots,0)\in
\boldsymbol{L},&& \mathcal{T}_9^{(\a_1+j)}\boldsymbol{g}:=
(g_1,\phi_{1,j})_{L_2(\Om_1^\b)}, && j=1,\ldots,p_1,
\\
&\boldsymbol{\phi}_{\a_2+j}:=(0,\phi_{2,j},0,\ldots,0)\in
\boldsymbol{L},&& \mathcal{T}_9^{(\a_2+j)}\boldsymbol{g}:=
(g_2,\phi_{2,j})_{L_2(\Om_2\b)}, && j=1,\ldots,p_2,
\\
&\hphantom{\boldsymbol{\phi}_{\a_2+j}:=(0,} \ldots
&&\hphantom{\mathcal{T}_9^{(q)}\boldsymbol{g}:}\ldots &&
\hphantom{j=} \ldots
\\
&\boldsymbol{\phi}_{\a_m+j}:=(0,0,\ldots,0,\phi_{m,j})\in
\boldsymbol{L},&& \mathcal{T}_9^{(\a_m+j)}\boldsymbol{g}:=
(g_m,\phi_{m,j})_{L_2(\Om_m^\b)}, && j=1,\ldots,p_m.
\end{align*}
Here $\phi_{i,j}:=\big(\mathcal{T}_1(\l_*)\big)^{-1}\psi_{i,j}$.
Lemmas~\ref{lm3.2},~\ref{lm4.1} yield

\begin{lemma}\label{lm4.2}
The operator $\mathcal{T}_8$ is bounded and holomorphic w.r.t.
$\l\in \mathbb{S}_\e$. For each compact set $\mathbb{K}\in
\mathbb{S}_\e$ the uniform in $\l\in\mathbb{K}$ and large
$l_{i,j}$ estimates
\begin{equation}\label{4.8}
\Big\|\frac{\p^i \mathcal{T}_8}{\p\l^i}\Big\|\leqslant C
l_X^{-\frac{n-2i-1}{2}}\E^{-l_X\sqrt{-\l}},\quad i=0,1,
\end{equation}
are valid. The operator $\mathcal{T}_7$ is bounded and
meromorphic w.r.t. $\l\in\mathbb{S}_\e$. The set of its poles
coincide with $\sigma_*$. For any $\l$ close to
$(p_1+\ldots+p_m)$-multiple $\l_*\in\sigma_*$ the representation
\begin{equation}
\big(\I+\mathcal{T}_7(\l)\big)^{-1}=-\sum\limits_{i=1}^{p}
\frac{\boldsymbol{\phi}_i
\mathcal{T}_9^{(i)}}{\l-\l_*}+\mathcal{T}_{10}(\l), \label{4.9}
\end{equation}
holds true, where the $j$-th component of the vector
$\mathcal{T}_{10}(\l)\boldsymbol{g}$ is $\mathcal{T}_3^{(j)}g_j$
if $p_j\not=0$ and
$\big(\I+\mathcal{T}_7^{(j)}(\l)\big)^{-1}g_j$ if $p_j=0$. The
operator $\mathcal{T}_{10}: \boldsymbol{L}\to \boldsymbol{L}$ is
bounded and holomorphic w.r.t. $\l$ close to $\l_*$. The
equation
$\big(\I+\mathcal{T}_7(\l_*)\big)\boldsymbol{g}=\boldsymbol{f}$
is solvable if and only if
$\mathcal{T}_9^{(i)}\boldsymbol{f}=0$, $i=1,\ldots,m$. The
solution of this equation is given by
$\boldsymbol{g}=\mathcal{T}_{10}(\l_*)\boldsymbol{f}+
\sum\limits_{i=1}^{p} c_i\boldsymbol{\phi}_i$, where $c_i$ are
some constants.
\end{lemma}

\begin{lemma}\label{lm4.3}
Each isolated eigenvalue of $\mathcal{H}_X$ converges to zero or
to $\l_*\in\sigma_*$ as $l_X\to+\infty$.
\end{lemma}

\begin{proof}
Using (\ref{1.2}), (\ref{1.3}), for each
$u\in\H^1(\mathbb{R}^n)$ we obtain
\begin{equation*}
(\mathcal{H}_Xu,u)_{L_2(\mathbb{R}^n)}\geqslant \|\nabla
u\|_{L_2(\mathbb{R}^n)}^2-c_0\|\nabla u\|_{L_2(\Om_X)}^2 -
c_1\|u\|_{L_2(\Om_X)}^2\geqslant -
c_1\|u\|_{L_2(\mathbb{R}^n)}^2,
\end{equation*}
which implies that 
$\discspec(\mathcal{H}_X)\subset[-c_1,0)$.  We define
$\mathbb{K}_\e:=[-c_1,-\e)\setminus\bigcup\limits_{\l\in\si_*}
(\l-\e,\l+\e)$. This set obeys the hypothesis of
Lemma~\ref{lm3.2}, and due to (\ref{4.8}) the norm of
$\mathcal{T}_8$ is exponentially small as $\l\in \mathbb{K}_\e$
and $l_X\to+\infty$. In accordance with Lemma~\ref{lm4.2}, the
operator $\I+\mathcal{T}_7(\l)$ is boundedly invertible as
$\l\in\mathbb{K}_\e$. Therefore, the operator
$\I+\mathcal{T}_7(\l)+T_8(\l,X)$ is boundedly invertible as
$\l\in \mathbb{K}_\e$ if $l_X$ is large enough. Thus, the
equation (\ref{4.7}) has no nontrivial solution as $\l\in
\mathbb{K}_\e$ if $l_X$ is large enough, and by
Lemma~\ref{lm3.1} we conclude that the set $\mathbb{K}_\e$
contains no eigenvalues of $\mathcal{H}_X$ if $l_X$ is large
enough. The number $\e$ being arbitrary completes the proof.
\end{proof}

Let $\l_*\in\sigma_*$ be $(p_1+\ldots+p_m)$-multiple; we are
going to find non-trivial solutions of (\ref{4.7}) for $\l$
close to $\l_*$.

Assume first that $\l\not=\l_*$. We apply the operator
$(\I+\mathcal{T}_7)^{-1}$ to this equation and then substitute
the representation (\ref{4.9}) into the relation obtained. This
procedure yields
\begin{equation}\label{4.16}
\boldsymbol{g}-\sum\limits_{i=1}^{p} \frac{\phi_i
\mathcal{T}_9^{(i)}\mathcal{T}_8(\l,X)\boldsymbol{g}}{\l-\l_*}
+\mathcal{T}_{10}(\l)\mathcal{T}_8(\l,X)\boldsymbol{g}=0.
\end{equation}
In view of  (\ref{4.8}) the operator
$\mathcal{T}_{10}(\l)\mathcal{T}_8(\l,X)$ is small if $l_X$ is
large enough. Thus, the operator
$\big(\I+\mathcal{T}_{10}(\l)\mathcal{T}_8(\l,X)\big)^{-1}$ is
well-defined and bounded. We apply now this operator to the
equation (\ref{4.16}) and arrive at
\begin{equation}\label{4.17}
\boldsymbol{g}-\sum\limits_{i=1}^{p}\frac{\mathcal{T}_9^{(i)}
\mathcal{T}_8(\l,X)\boldsymbol{g}}{\l-\l_*}
\boldsymbol{\Phi}_i=0,
\end{equation}
where $\boldsymbol{\Phi}_i(\cdot,\l,X):=
\big(\I+\mathcal{T}_{10}(\l)\mathcal{T}_8(\l,X)\big)^{-1}
\boldsymbol{\phi}_i$. Hence,
\begin{equation}\label{4.10}
\boldsymbol{g}=\sum\limits_{i=1}^{p} \kappa_i
\boldsymbol{\Phi}_i,
\end{equation}
where $\kappa_i$ are some numbers to be found. We substitute now
this identity into (\ref{4.17}) and obtain
\begin{gather}
\sum\limits_{i=1}^{p}\boldsymbol{\Phi}_i\left(\kappa_i-
\sum\limits_{j=1}^{p}A_{ij}\kappa_j\right)=0,\label{4.18}
\\
A_{ij}=A_{ij}(\l,X):=\mathcal{T}_9^{(i)}
\mathcal{T}_8(\l,X)\boldsymbol{\Phi}_j(\cdot,\l,X).\nonumber
\end{gather}
The estimates (\ref{4.8}) imply that
\begin{equation}\label{4.19}
\boldsymbol{\Phi}_i=\boldsymbol{\phi}_i+
\Odr(l_X^{-\frac{n-1}{2}}\E^{-l_X\sqrt{-\l}}),\quad
l_X\to+\infty.
\end{equation}
Since the vectors $\boldsymbol{\phi}_i$ are linear independent,
due to these relations the same is true for
$\boldsymbol{\Phi}_i$. Thus, the equation (\ref{4.18}) is
equivalent to the system of linear equations
\begin{gather}
\big((\l-\l_*)\mathrm{E}-\mathrm{A}(\l,X)\big)\boldsymbol{\kappa}=0,
\label{4.11}
\\
\boldsymbol{\kappa}:=
\begin{pmatrix}
\kappa_1
\\
\vdots
\\
\kappa_p
\end{pmatrix},
\quad \mathrm{A}(\l,X):=
\begin{pmatrix}
A_{11}(\l,X)& \ldots& A_{1p}(\l,X)
\\
\vdots& &\vdots
\\
A_{p1}(\l,X)&\ldots&A_{pp}(\l,X)
\end{pmatrix},\label{4.12}
\end{gather}
where $\mathrm{E}$ is the identity matrix. The corresponding
solution of the equation (\ref{4.7}) is given by (\ref{4.10}).
Since the vectors $\boldsymbol{\Phi}_i$ are linear independent,
this solution is non-zero if and only if
$\boldsymbol{\kappa}\not=0$. The criterion of the existence of
nontrivial solution to (\ref{4.11}) is
\begin{equation}\label{4.15}
\det\big((\l-\l_*)\mathrm{E}-\mathrm{A}(\l,X)\big)=0.
\end{equation}
Therefore, the number $\l\not=\l_*$ converging to $\l_*$ as
$l_X\to+\infty$ is an eigenvalue the operator $\mathcal{H}_X$ if
and only if it is a root of the obtained equation. The
multiplicity of this eigenvalue equals to the number of linear
independent solutions of the corresponding system (\ref{4.11}).
Let us prove that the same is true if $\l=\l_*$.

Consider the equation (\ref{4.7}) with $\l=\l_*$. If we treat
$\mathcal{T}_8(\l_*,X)\boldsymbol{g}$ as a right-hand side,
according to Lemma~\ref{lm4.2} this equation is solvable if and
only if
\begin{equation}\label{4.14}
\mathcal{T}_9^{(i)} \mathcal{T}_8(\l_*,X) \boldsymbol{g}=0,
\quad i=1,\ldots,m,
\end{equation}
and the solution is given by
\begin{equation*}
\boldsymbol{g}=-\mathcal{T}_{10}(\l_*)\mathcal{T}_8(\l_*,X)g+
\sum\limits_{i=1}^{p} \kappa_i\boldsymbol{\phi}_i,
\end{equation*}
where $\kappa_i$ are some constants.  Now we apply the operator
$\big(\I+\mathcal{T}_{10}(\l_*)\mathcal{T}_8(\l_*,X)\big)^{-1}$
to this identity and arrive at the formula (\ref{4.10}) with
$\l=\l_*$. We substitute this formula into (\ref{4.14}) and
obtain the system (\ref{4.11}) with $\l=\l_*$. The vector
$\boldsymbol{g}$ is non-zero if and only if
$\boldsymbol{\kappa}\not=0$; this leads us to the equation
(\ref{4.15}) with $\l=\l_*$.

It is convenient to summarize the obtained results in
\begin{lemma}\label{lm4.4}
Let $\l_*\in\sigma_*$ be $(p_1+\ldots+p_m)$-multiple. A number
$\l\xrightarrow[l_X\to+\infty]{}\l_*$ is an eigenvalue the
$\mathcal{H}_X$ if and only if it is a root of (\ref{4.15}). The
multiplicity of this eigenvalue equals to the number of linear
independent solutions of the corresponding system (\ref{4.11}).
\end{lemma}


\section{Proof of Theorems~\ref{th1.2}--\ref{th1.4}}

In view of Lemmas~\ref{lm4.3},~\ref{lm4.4} we will complete the
proof of Theorem~\ref{th1.2}, if we prove that total number of
non-trivial solutions to (\ref{4.11}) associated with the roots
of (\ref{4.15}) equals $p$.

Throughout this section we assume that $\l_*\in\sigma_*$ is
$(p_1+\ldots+p_m)$-multiple and $\l$ belongs to a small
neighbourhood of $\l_*$. We denote
$\mathrm{B}(\l,X):=(\l-\l_*)\mathrm{E}-\mathrm{A}(\l,X)$,
$F(\l,X):=\det\mathrm{B}(\l,X)$.

\begin{lemma}\label{lm5.0}
In the vicinity of $\l_*$ the function $F(\l,X)$ has exactly $p$
zeroes counting their orders. These zeroes converge to $\l_*$ as
$l_X\to+\infty$.
\end{lemma}
\begin{proof}
The definition of the functions $A_{i,j}$ and Lemma~\ref{lm4.2}
imply that these functions are holomorphic w.r.t. $\l$ and
satisfy the estimates
\begin{equation*}
\left|\frac{\p^k A_{ij}}{\p\l^k}(\l,l)\right|\leqslant C
l_X^{-\frac{n-2k-1}{2}}\E^{-l_X\sqrt{-\l}},\quad k=0,1.
\end{equation*}
It is clear that
\begin{equation*}
F(\l,X)=(\l-\l_*)^p+\sum\limits_{i=0}^{p-1}P_i(\l,X)(\l-\l_*)^i,
\end{equation*}
where the functions $P_i$ are holomorphic w.r.t. $\l$ and obey
the uniform in  estimate
\begin{equation*}
|P_i(\l,X)|\leqslant
Cl_X^{-\frac{(p-i)(n-1)}{2}}\E^{-(p-i)l_X\sqrt{-\l}}.
\end{equation*}
For a sufficiently small fixed $\e>0$ this estimate yields
\begin{equation*}
\left|\sum\limits_{i=0}^{p-1}P_i(\l,X)(\l-\l_*)^i\right|
<|\l-\l_*|^p \quad \text{as}\quad |\l-\l_*|=\e,
\end{equation*}
if $l_X$ is large enough. Hence, by Rouche theorem the function
$F(\l,X)$ has the same number of zeroes (counting orders) inside
the disk $\{\l: |\l-\l_*|<\e\}$ as the function $\l\mapsto
(\l-\l_*)^p$ does. The number $\e$ being arbitrary completes the
proof.
\end{proof}

\begin{lemma}\label{lm5.1}
Suppose that $\l_1(X)$ and $\l_2(X)$ are different roots of the
equation (\ref{4.15}), and $\boldsymbol{\kappa}_1(X)$ and
$\boldsymbol{\kappa}_2(X)$ are the associated non-trivial
solutions to the system (\ref{4.11}) normalized by the condition
\begin{equation*}
\|\boldsymbol{\kappa}_i\|_{\mathbb{C}^p}=1.
\end{equation*}
Then
\begin{equation*}
\big(\boldsymbol{\kappa}_1,\boldsymbol{\kappa}_2\big)_{\mathbb{C}^p}=
\Odr(l_X^{\frac{n-1}{2}}\E^{-l_X\sqrt{-\l}}),\quad
l_X\to+\infty.
\end{equation*}
\end{lemma}

\begin{proof}
We indicate by $\boldsymbol{g}_j$ the solutions of the equation
(\ref{4.7}) associated with $\l_j$; these solutions are given by
(\ref{4.10}). Due to Lemma~\ref{lm3.4} the functions
$\widetilde{v}_i$ and $\widetilde{w}_i$ corresponding to each of
the vectors $\boldsymbol{g}_j$ satisfy the estimates
\begin{equation*}
\|\mathcal{L}_i\widetilde{v}_i\|_{L_2(\Om_i^\b)}=
\Odr(l_X^{-\frac{(n-1)}{2}}\E^{-l_X\sqrt{-\l_j}}),\quad
\|\widetilde{w}_i\|_{\H^2(\Om_i^\b)}=
\Odr(l_X^{-\frac{(n-1)}{2}}\E^{-l_X\sqrt{-\l_j}}),
\end{equation*}
as $l_X\to+\infty$. Moreover, if follows from (\ref{4.19}) that
\begin{equation*}
\boldsymbol{g}_i=\sum\limits_{j=1}^{p} \kappa_j^{(i)}
\boldsymbol{\phi}_j
+\Odr(l_X^{-\frac{n-1}{2}}\E^{-l_X\sqrt{-\l_i}}),
\end{equation*}
where $\kappa_j^{(i)}$ are the components of the vectors
$\boldsymbol{\kappa}_i$. In view of the relation obtained and
(\ref{4.1}), (\ref{4.2}) we infer that the eigenfunctions
$\psi_i(x,X)$ associated with $\l_i$ satisfy the asymptotic
formulas:
\begin{equation*}
\psi_i=\sum\limits_{j=1}^{m} \mathcal{S}(-X_j)
\sum\limits_{q=1}^{p_j} \kappa_{\a_j+q}^{(i)} \psi_{j,q}
+\Odr\big(l_X^{-\frac{n-1}{2}}\E^{-l_X\sqrt{-\l_j}}\big),
\end{equation*}
where, we remind,  $\psi_{i,j}$, $j=1,\ldots,p_i$, are the
eigenfunctions of $\mathcal{H}_i$ associated with $\l_*$ and
orthonormalized in $L_2(\mathbb{R}^n)$. Since the operator
$\mathcal{H}_{X}$ is self-adjoint, the eigenfunctions $\psi_i$
are orthogonal in $L_2(\mathbb{R}^n)$. Together with the
established asymptotic representations for $\psi_i$ it implies
\begin{equation}\label{5.4}
\begin{aligned}
0=(\psi_1,\psi_2)_{L_2(\mathbb{R}^n)}=&\sum\limits_{i,j=1}^{m}
\sum\limits_{q=1}^{p_j}\sum\limits_{r=1}^{p_i}
\kappa_{\a_j+q}^{(1)} \overline{\kappa}_{\a_i+r}^{(2)}
\Big(\mathcal{S}(-X_j) \psi_{j,q}, \mathcal{S}(-X_i)
\psi_{i,r}\Big)_{L_2(\mathbb{R}^n)}
\\
&+
\Odr\big(l_X^{-\frac{n-1}{2}}\E^{-l_X\sqrt{-\max\{\l_1,\l_2\}}}
\big).
\end{aligned}
\end{equation}
It is clear that
\begin{equation*}
\Big(\mathcal{S}(-X_j) \psi_{j,q}, \mathcal{S}(-X_j)
\psi_{j,r}\Big)_{L_2(\mathbb{R}^n)}= (\psi_{j,q},
\psi_{j,r})_{L_2(\mathbb{R}^n)}=
\begin{cases}
1, & q=r,
\\
0, & q\not=r,
\end{cases}
\end{equation*}
and for $i\not=j$
\begin{align*}
&\Big(\mathcal{S}(-X_j) \psi_{j,q}, \mathcal{S}(-X_i)
\psi_{i,r}\Big)_{L_2(\mathbb{R}^n)}=\Big(\mathcal{S}(X_{i,j})
\psi_{j,q}, \psi_{i,r}\Big)_{L_2(\mathbb{R}^n)}
\\
&=\Big(\mathcal{S}(X_{i,j}) \psi_{j,q},
\psi_{i,r}\Big)_{L_2\left(\Om_i^{l_{i,j}/2}\right)}+
\Big(\mathcal{S}(X_{i,j}) \psi_{j,q},
\psi_{i,r}\Big)_{L_2\left(\mathbb{R}^n
\setminus\Om_i^{l_{i,j}/2}\right)}=\Odr\left(l_{i,j}^{\frac{n-1}{2}}
\E^{-l_{i,j}\sqrt{-\l_*}}\right),
\end{align*}
Here we have used that due to (\ref{3.16})
\begin{equation*}
\psi_{i,j}=C_{i,j}|x|^{-(n-1)/2}\E^{-|x|\sqrt{-\l}}
\Big(1+\Odr\big(|x|^{-1}\big)\Big),\quad |x|\to+\infty,
\end{equation*}
where $C_{i,j}$ are some constants. Substituting the obtained
relations into (\ref{5.4}), we arrive at the statement of the
lemma.
\end{proof}

Let $\l(X)\xrightarrow[l_X\to+\infty]{}\l_*$ be a root of the
equation (\ref{4.15}). Without loss of generality we assume that
the corresponding solutions of (\ref{4.11}) are orthonormalized
in $\mathbb{C}^p$. Consider the set of all such solutions to
(\ref{4.11}) associated with all roots of (\ref{4.15})
converging to $\l_*$ as $l_X\to+\infty$, and denote these
vectors as $\boldsymbol{\kappa}_i=\boldsymbol{\kappa}_i(X)$,
$i=1,\ldots,q$. In view of Lemma~\ref{lm5.1} the vectors
$\boldsymbol{\kappa}_i$ satisfy the formulas (\ref{1.4}).

\begin{lemma}\label{lm5.2}
Let $\l(X)\xrightarrow[l_X\to+\infty]{}\l_*$ be a root of the
equation (\ref{4.15}) and $\boldsymbol{\kappa}_i$,
$i=N,\ldots,N+q$, $q\geqslant0$, be the associated solutions to
(\ref{4.11}). Then the representation
\begin{equation*}
\mathrm{B}^{-1}(\l,X)=\sum\limits_{i=N}^{N+q}
\frac{\mathcal{T}_{11}^{(i)}(X)}{\l-\l(X)}
\boldsymbol{\kappa}_i(X)+\mathrm{B}_0(\l,X)
\end{equation*}
is valid for all $\l$ close to $\l(X)$. Here
$\mathcal{T}_{11}^{(i)}:\mathbb{C}^p\to\mathbb{C}$ are some
functionals, while the matrix $\mathrm{B}_0(\l,X)$ is
holomorphic w.r.t. $\l$ in a neighbourhood of $\l(X)$.
\end{lemma}

\begin{proof}
The matrix $\mathrm{B}$ is meromorphic and its inverse thus has
a pole at $\l(X)$. By analogy with the relations (5.7), (5.8) in
\cite{B1} one can show that the residue at this pole is of the
form
$\sum_{i=N}^{N+q}\boldsymbol{\kappa}_i(X)\mathcal{T}_{11}^{(i)}(X)$,
where $\mathcal{T}_{11}^{(i)}: \mathbb{C}^p\to \mathbb{C}$ are
some functionals. We are going to prove that this pole is
simple; clearly, it will complete the proof of the lemma.

Consider $\l$ close to $\l(X)$ and not coinciding with $\l_*$
and $\l(X)$. Let $f_i\in L_2(\mathbb{R}^n;\Om_i)$ be arbitrary
functions, $\boldsymbol{f}:=(f_1,\ldots,f_m)\in\boldsymbol{L}$,
$\widetilde{f}:=\sum_{i=1}^{m} \mathcal{S}(-X_i)f_i$. Completely
by analogy with (\ref{4.1})--(\ref{4.7}) one can check easily
that the equation (\ref{3.4}) with
$\mathcal{T}_2=\mathcal{T}_2^{(X)}$ is equivalent to
\begin{equation*}
\boldsymbol{g}+\mathcal{T}_7(\l)\boldsymbol{g}+
\mathcal{T}_8(\l,X)\boldsymbol{g}=\boldsymbol{f}.
\end{equation*}
Proceeding as in (\ref{4.16}), (\ref{4.17}), one can reduce this
equation to an equivalent one
\begin{equation}\label{5.3}
\boldsymbol{g}-\sum\limits_{i=1}^{p}\frac{\mathcal{T}_9^{(i)}
\mathcal{T}_8(\l,X)\boldsymbol{g}}{\l-\l_*}\boldsymbol{\Phi}_i=
-\sum\limits_{i=1}^{p}\frac{\mathcal{T}_9^{(i)}
\boldsymbol{f}}{\l-\l_*}\boldsymbol{\Phi}_i
+\big(\I+\mathcal{T}_{10}(\l)\mathcal{T}_8(\l,X)\big)^{-1}
\mathcal{T}_{10}(\l)\boldsymbol{f}.
\end{equation}
We denote
\begin{equation*}
\kappa_i:=\frac{\mathcal{T}_9^{(i)}
\mathcal{T}_8(\l,X)\boldsymbol{g}}{\l-\l_*}
\end{equation*}
and apply the functionals $\mathcal{T}_9^{(j)}
\mathcal{T}_8(\l,X)$ to (\ref{5.3}). This procedure leads us to
the equation for $\kappa_i$:
\begin{gather}
\mathrm{B}(\l,X)\boldsymbol{\kappa}=-\frac{1}{\l-\l_*}
\mathrm{A}(\l,X)\boldsymbol{h}_1+\boldsymbol{h}_2, \nonumber
\\
\boldsymbol{h}_1
:=
\begin{pmatrix}
\mathcal{T}_9^{(1)}\boldsymbol{f}
\\
\vdots
\\
\mathcal{T}_9^{(p)}\boldsymbol{f}
\end{pmatrix},
\quad \boldsymbol{h}_2
:=
\begin{pmatrix}
\mathcal{T}_9^{(1)}\mathcal{T}_8(\l,X)
\big(\I+\mathcal{T}_{10}(\l)\mathcal{T}_8(\l,X)\big)^{-1}
\mathcal{T}_{10}(\l)\boldsymbol{f}
\\
\vdots
\\
\mathcal{T}_9^{(p)}\mathcal{T}_8(\l,X)
\big(\I+\mathcal{T}_{10}(\l)\mathcal{T}_8(\l,X)\big)^{-1}
\mathcal{T}_{10}(\l)\boldsymbol{f}
\end{pmatrix},\label{5.6}
\end{gather}
where $\boldsymbol{\kappa}$ is defined as in (\ref{4.12}).
Hence,
\begin{gather*}
\boldsymbol{\kappa}=\frac{1}{\l-\l_*}\boldsymbol{h}_1+
\boldsymbol{\widetilde{\kappa}},\quad
\boldsymbol{\widetilde{\kappa}}:=
\mathrm{B}^{-1}\boldsymbol{\widetilde{h}}, \quad
\boldsymbol{\widetilde{h}}:= \boldsymbol{h}_2-\boldsymbol{h}_1,
\\
\boldsymbol{g}=\sum\limits_{i=1}^{p}
\widetilde{\kappa}_i\boldsymbol{\Phi}_i+
\big(\I+\mathcal{T}_{10}(\l)\mathcal{T}_8(\l,X)\big)^{-1}
\mathcal{T}_{10}(\l)\boldsymbol{f},
\end{gather*}
where $\widetilde{\kappa}_i$ are components of the vector
$\widetilde{\boldsymbol{\kappa}}$. In accordance with
Lemma~\ref{lm3.2} the solution to the equation (\ref{3.4}) with
$\mathcal{T}_2=\mathcal{T}_2^{(X)}$ has at most simple pole at
$\l(X)$. Hence, the same is true for the vector $\boldsymbol{g}$
just determined. It follows that the vector
$\mathrm{B}^{-1}\boldsymbol{\widetilde{h}}$ can have at most
simple pole at $\l(X)$. The estimates (\ref{4.8}) imply that
\begin{equation*}
\boldsymbol{\widetilde{h}}=-\boldsymbol{h}_1+\Odr\big(
l_X^{-\frac{n-1}{2}}\E^{-l_X\sqrt{-\l}} \big).
\end{equation*}
In view of this identity and the definition of
$\boldsymbol{h}_1$ we conclude that for any
$\widetilde{\boldsymbol{h}}\in \mathbb{C}^p$ there exists
$\boldsymbol{f}\in\boldsymbol{L}$ such that
$\widetilde{\boldsymbol{h}}=\boldsymbol{h}_2-\boldsymbol{h}_1$,
where $\boldsymbol{h}_i$ are given by (\ref{5.6}). Therefore,
the matrix $\mathrm{B}^{-1}$ has the simple pole at $\l(X)$.
\end{proof}

Reproducing word for word the proof of Lemma~5.3 in \cite{B1} we
obtain

\begin{lemma}\label{lm5.3}
A zero $\l(X)\xrightarrow[l_X\to+\infty]{}\l_*$ of the function
$F(\l,X)$ has order $q$ if and only if it is a $q$-multiple
eigenvalue of $\mathcal{H}_X$.
\end{lemma}

The statement of Theorem~\ref{th1.2} follows from
Lemmas~\ref{lm4.3},~\ref{lm4.4},~\ref{lm5.0},~\ref{lm5.3}.

The proof of Theorems~\ref{th1.3},~\ref{th1.4} repeats
\emph{verbatim et literatim} the proof of Theorems~1.4,~1.5 in
\cite{B1}.


\section{Proof of Theorems~\ref{th1.5},~\ref{th1.7} and
Corollary~\ref{cr1.6}}

\begin{proof}[Proof of Theorem~\ref{th1.5}.] Let us prove first
that the representation (\ref{1.9}) is valid, where the matrix
$\mathrm{A}_0$ is defined in the statement of the theorem and
\begin{equation*}
\|\mathrm{A}_1\|=\Odr\left(l_X^{-n+1}
\E^{-2l_X\sqrt{-\l_*}}\right),\quad l_X\to+\infty.
\end{equation*}
Due to (\ref{4.8}), (\ref{4.19}) we have
\begin{equation*}
A_{i,j}(\l_*,X)=\mathcal{T}_9^{(i)}
\mathcal{T}_8(\l_*,X)\boldsymbol{\phi}_j+\Odr\left(l_X^{-n+1}
\E^{-2l_X\sqrt{-\l_*}}\right),\quad l_X\to+\infty.
\end{equation*}
We are going to show that $A_{i,j}^{(0)}(X)=\mathcal{T}_9^{(i)}
\mathcal{T}_8(\l_*,X)\boldsymbol{\phi}_j$ and the matrix
$\mathrm{A}_0$ satisfies the condition (A); this will obviously
imply the needed representation.

We choose some $i$ and $j$ and let $k,r\in\{1,\ldots,m\}$,
$q\in\{1,\ldots,p_k\}$, $s\in\{1,\ldots,p_r\}$ be such that
$i=\a_k+q$, $j=\a_r+s$. Then
\begin{equation*}
\mathcal{T}_9^{(i)} \mathcal{T}_8(\l_*,X)\boldsymbol{\phi}_j=
\begin{cases}
\big(\mathcal{T}_6^{(k,r)}(\l_*)\phi_{r,s},
\psi_{k,q}\big)_{L_2(\Om_k)},& r\not=k,
\\
\hphantom{\big(\mathcal{T}_6^{(k,r)}(\l_*)}0,& r=k.
\end{cases}
\end{equation*}
Consider the case $r\not=k$. We employ (\ref{4.5a}) and
(\ref{1.1}) and integrate by parts to obtain:
\begin{equation}\label{6.1}
\begin{aligned}
\big(\mathcal{T}_6^{(k,r)}&(\l_*)\phi_{r,s},
\psi_{k,q}\big)_{L_2(\Om_k)}=\big(\mathcal{L}_k
\mathcal{T}_5(\l_*,X_{k,r})\phi_{r,s},
\psi_{k,q}\big)_{L_2(\Om_k)}
\\
&+\big((\D-\l_*+\mathcal{L}_k)\chi_{\Om_k}
(\mathcal{H}_{\Om_k}-\iu)^{-1}\mathcal{L}_k
\mathcal{T}_5(\l,X_{k,r})\phi_{r,s},\psi_{k,q}\big)_{L_2(\Om_k)}
\\
&= \big(\mathcal{L}_k \mathcal{T}_5(\l_*,X_{k,r})\phi_{r,s},
\psi_{k,q}\big)_{L_2(\Om_k)}.
\end{aligned}
\end{equation}
It follows from Lemma~\ref{lm3.1} and the definition of
$\mathcal{T}_5$ that $\mathcal{T}_5(\l_*,X_{k,r})\phi_{k,s}=
\mathcal{S}(X_{k,r})\psi_{r,s}$. Hence,
\begin{equation*}
\mathcal{T}_9^{(i)}\mathcal{T}_8(\l_*,X)\boldsymbol{\phi}_j=
\big(\mathcal{L}_k\mathcal{S}(X_{k,r})\psi_{r,s},
\psi_{k,q}\big)_{L_2(\Om_k)}= A_{i,j}^{(0)}(X).
\end{equation*}
Using this identity, the condition (\ref{1.1}) and the equation
for $\psi_{r,s}$ and $\psi_{k,q}$, we check that
\begin{equation}
\begin{aligned}
A_{i,j}^{(0)}(X)&=
\big(\mathcal{L}_k\mathcal{S}(X_{k,r})\psi_{r,s},
\psi_{k,q}\big)_{L_2(\Om_k)}=
\big(\mathcal{S}(X_{k,r})\psi_{r,s},
\mathcal{L}_k\psi_{k,q}\big)_{L_2(\Om_k)}
\\
&=\big(\psi_{r,s}, \mathcal{S}(X_{r,k})\mathcal{L}_k
\psi_{k,q}\big)_{L_2(\mathbb{R}^n)}= \big(\psi_{r,s}, (\D+\l_*)
\mathcal{S}(X_{r,k})\psi_{k,q}\big)_{L_2(\mathbb{R}^n)}
\\
&=\big((\D+\l_*)\psi_{r,s},
\mathcal{S}(X_{r,k})\psi_{k,q}\big)_{L_2(\mathbb{R}^n)}=
\big(\mathcal{L}_r\psi_{r,s},
\mathcal{S}(X_{r,k})\psi_{k,q}\big)_{L_2(\Om_r)}
\\
&=\overline{\big(\mathcal{L}_r\mathcal{S}(X_{r,k})\psi_{k,q},
\psi_{r,s}\big)}_{L_2(\Om_r)}= \overline{A}_{j,i}^{(0)}(X).
\end{aligned}\label{6.1a}
\end{equation}
Hence, the matrix $\mathrm{A}_0$ is hermitian. The eigenvectors
of $\mathrm{A}_0$ are orthonormal in $\mathbb{C}^p$, and the
determinant of the matrix formed by these vectors thus equals
one. Therefore, the matrix $\mathrm{A}_0$ satisfies the
condition (A). Now it is sufficient to apply Theorem~\ref{th1.4}
to complete the proof.
\end{proof}

\begin{proof}[Proof of Corollary~\ref{cr1.6}]
In the case considered the matrix $\mathrm{A}_0$ reads as
follows:
\begin{equation*}
\mathrm{A}_0=
\begin{pmatrix}
0 & \big(\mathcal{L}_1
\mathcal{S}(X_{1,2})\psi_2,\psi_1\big)_{L_2(\Om_1)}
\\
\overline{\big(\mathcal{L}_1
\mathcal{S}(X_{1,2})\psi_2,\psi_1\big)}_{L_2(\Om_1)} & 0
\end{pmatrix},
\end{equation*}
where we have taken in account the hermiticity of this matrix.
The eigenvalues of $\mathrm{A}_0$ are
$\tau_1^{(0)}=-\left|\big(\mathcal{L}_1
\mathcal{S}(X_{1,2})\psi_2,\psi_1\big)_{L_2(\Om_1)}\right|$,
$\tau_2^{(0)}=\left|\big(\mathcal{L}_1
\mathcal{S}(X_{1,2})\psi_2,\psi_1\big)_{L_2(\Om_1)}\right|$.
Applying now Theorem~\ref{th1.5}, we complete the proof.
\end{proof}

\begin{proof}[Proof of Theorem~\ref{th1.7}]
Theorem~\ref{th1.3} implies that the eigenvalue $\l(X)$ has the
asymptotic expansion (\ref{1.6}), where
$\tau(X)=A_{11}(\l_*,X)$. It follows from the definition of
$\boldsymbol{\Phi}_1$ and the estimates (\ref{4.8}) that
\begin{equation*}
\boldsymbol{\Phi}_1(\cdot,\l_*,X)=\boldsymbol{\phi}_1-
\mathcal{T}_{10}(\l_*)\mathcal{T}_8(\l_*,X)\boldsymbol{\phi}_1
+\Odr\big(l_X^{-n+1}\E^{-2l_X\sqrt{-\l_*}}\big),\quad
l_X\to+\infty.
\end{equation*}
Since $\mathcal{T}_9^{(1)}\mathcal{T}_8(\l_*,X)
\boldsymbol{\phi}_1=0$, we infer that
\begin{equation}\label{6.2}
A_{11}(\l_*,X)=-\mathcal{T}_9^{(1)}\mathcal{T}_8(\l_*,X)
\mathcal{T}_{10}(\l_*)\mathcal{T}_8(\l_*,X)\boldsymbol{\phi}_1
+\Odr\big(l_X^{-\frac{3n-3}{2}} \E^{-3l_X\sqrt{-\l_*}}\big),
\end{equation}
as $l_X\to+\infty$. By direct calculations we check that
\begin{equation*}
\mathcal{T}_{10}(\l_*)\mathcal{T}_8(\l_*,X)\boldsymbol{\phi}_1 =
\Big(0,\big(\I+\mathcal{T}_7^{(2)}(\l_*)\big)^{-1}
\mathcal{T}_6^{(2,1)}\phi_1,\ldots,
\big(\I+\mathcal{T}_7^{(m)}(\l_*)\big)^{-1}
\mathcal{T}_6^{(m,1)}\phi_1\Big),
\end{equation*}
where $\phi_1:=\big(\mathcal{T}_1^{(1)}(\l_*)\big)^{-1}\psi_1$.
Using this relation and proceeding in the same way as in
(\ref{6.1}), we obtain
\begin{equation}\label{6.3}
\begin{aligned}
\mathcal{T}_9^{(1)}&\mathcal{T}_8(\l_*,X)
\mathcal{T}_{10}(\l_*)\mathcal{T}_8(\l_*,X)\boldsymbol{\phi}_1
\\
&=\sum\limits_{j=2}^{m} \Big( \mathcal{L}_1
\mathcal{S}(X_{1,j})\mathcal{T}_1^{(j)}(\l_*)
\big(\I+\mathcal{T}_7^{(j)}(\l_*)\big)^{-1}\mathcal{T}_6^{(j,1)}
\phi_1,\psi_1\Big)_{L_2(\Om_1)}.
\end{aligned}
\end{equation}
In accordance with Lemma~\ref{lm3.1} the function
$\mathcal{T}_1^{(j)}(\l_*)
\big(\I+\mathcal{T}_7^{(j)}(\l_*)\big)^{-1}\mathcal{T}_6^{(j,1)}
\phi_1$, $j=2,\ldots,m$, is a solution to the equation
(\ref{3.1}) with $\mathcal{H}_{\mathcal{L}}=\mathcal{H}_j$,
$\l=\l_*$, $f=\mathcal{T}_6^{(j,1)}\phi_1$.  Since
\begin{equation*}
\mathcal{T}_6^{(j,1)}\phi_1=\mathcal{L}_j
\mathcal{T}_5(\l_*,X_{j,1})\phi_1
-(\mathcal{H}_j-\l_*)\chi_{\Om_j} (\mathcal{H}_{\Om_j}-\iu)^{-1}
\mathcal{L}_j \mathcal{T}_5(\l_*,X_{j,1})\phi_1,
\end{equation*}
due to (\ref{4.5a}), and
$\mathcal{L}_j\mathcal{T}_5(\l_*,X_{j,1})\phi_1=\mathcal{L}_j
\mathcal{S}(X_{j,1})\psi_1$ by Lemma~\ref{lm3.1}, we infer that
\begin{align*}
\mathcal{T}_1^{(j)}(\l_*)
\big(\I+\mathcal{T}_7^{(j)}(\l_*)\big)^{-1}\mathcal{T}_6^{(j,1)}
\phi_1=&(\mathcal{H}_j-\l_*)^{-1}\mathcal{L}_j
\mathcal{S}(X_{j,1})\psi_1
\\
&- \chi_{\Om_j} (\mathcal{H}_{\Om_j}-\iu)^{-1} \mathcal{L}_j
\mathcal{S}(X_{j,1})\psi_1.
\end{align*}
The support of the second term in the right-hand side of the
obtained identity lies inside $\Om_j^\b$. Bearing this fact in
mind, from (\ref{6.3}) we deduce
\begin{align*}
\mathcal{T}_9^{(1)}\mathcal{T}_8(\l_*,X)
\mathcal{T}_{10}(\l_*)&\mathcal{T}_8(\l_*,X)\boldsymbol{\phi}_1
\\
&=\sum\limits_{j=2}^{m} \Big( \mathcal{L}_1
\mathcal{S}(X_{1,j})(\mathcal{H}_j-\l_*)^{-1}\mathcal{L}_j
\mathcal{S}(X_{j,1})\psi_1,\psi_1\Big)_{L_2(\Om_1)}.
\end{align*}
We substitute this identity and (\ref{6.2}) into (\ref{1.6}) and
take into account that by (\ref{4.8})
\begin{equation*}
\mathcal{T}_9^{(1)}\mathcal{T}_8(\l_*,X)
\mathcal{T}_{10}(\l_*)\mathcal{T}_8(\l_*,X)\boldsymbol{\phi}_1=
\Odr\big(l_X^{-n+1}\E^{-2l_X\sqrt{-\l_*}}\big),\quad
l_X\to+\infty.
\end{equation*}
This leads us to the claimed asymptotics for $\l(X)$.

Since $p=1$, the system (\ref{4.11}) reduces to an equation
$(\l-\l_*-A_{11}(\l,X))\k_1=0$, which has the non-trivial
solution $\k_1=1$. This identity and Theorem~\ref{th1.3} imply
the asymptotics for $\psi(x,X)$.
\end{proof}


\section{Examples}

In this section we will give some possible examples of the
operators $\mathcal{L}_i$. Throughout this section we suppose
that $\Om_i\subset \mathbb{R}^n$ are given bounded domains with
infinitely differentiable boundary. We will often omit the index
''$i$'' in the notations corresponding to $i$-th operator
$\mathcal{L}_i$ writing simply $\mathcal{L}$, $\Om$,
$\mathcal{H}$, etc.

\textbf{1. Potential.} The simplest example of the operator
$\mathcal{L}$ is the multiplication by the compactly supported
real-valued potential. This is a classical example but it seems
that in the multiple-well case $m\geqslant 3$ the asymptotics
expansions for the eigenvalues were not known.

\textbf{2. Second order differential operator.} A more general
example is a differential operator of the form
\begin{equation}\label{7.1}
\mathcal{L}=\sum\limits_{i,j=1}^{n} b_{ij}\frac{\p^2}{\p x_i\p
x_j}+ \sum\limits_{i=1}^{n} b_{i}\frac{\p}{\p x_i}+b_0,
\end{equation}
where the coefficients $b_{ij}$ are piecewise continuously
differentiable and the coefficients $b_i$ are piecewise
continuous. The functions $b_{ij}$ and $b_i$ are also assumed to
be complex-valued and compactly supported. We also suppose that
the conditions (\ref{1.1}), (\ref{1.2}) hold true; the
self-adjointness of the operator $\mathcal{H}$ and
$\mathcal{H}_X$ follows from these conditions due to specific
definition of $\mathcal{L}$.

The particular case of (\ref{7.1}) is
\begin{equation}\label{7.2}
\mathcal{L}=\dvr \mathrm{G}\nabla+\iu\sum\limits_{i=1}^{n}
\left(b_i\frac{\p}{\p x_i}-\frac{\p}{\p x_i}b_i\right)+b_0,
\end{equation}
where $\mathrm{G}=\mathrm{G}(x)$ is $n\times n$ hermitian matrix
having piecewise continuously differentiable elements, the
functions $b_i=b_i(x)$ are real-valued and piecewise
continuously differentiable, the potential $b_0=b_0(x)$ is a
real-valued and piecewise continuous. We also suppose that the
matrix $\mathrm{G}$ and the functions $b_i$ are compactly
supported and
\begin{equation*}
|(\mathrm{G}(x)y,y)_{\mathbb{C}^n}|\leqslant
c_0\|y\|^2_{\mathbb{C}^n}, \qquad x\in\overline{\Pi},\quad
y\in\mathbb{C}^n,
\end{equation*}
where the constant $c_0$ is independent of $x$, $y$ and obeys
(\ref{1.3}). The matrix $\mathrm{G}$ can be zero; in the case
the operator $\mathcal{L}$ is a first order differential
operator.

\textbf{3. Magnetic Schr\"odinger operator.} Let
$\boldsymbol{b}=(b_1,\ldots,b_n)\in C_0^1(\mathbb{R}^n)$ be a
magnetic vector-potential, and
$b_0:=\|\boldsymbol{b}\|_{\mathbb{R}^n}+V$, where $V=V(x)\in
C_0^\infty(\mathbb{R}^n)$ is an electric potential. We define
the operator $\mathcal{L}$ by the formula (\ref{7.2}) with
$G=0$. Such operator describes the magnetic field with compactly
supported vector-potential.

\textbf{4. Integral operator.} The operator $\mathcal{L}$ is not
necessary to be a differential one. For instance, it can be an
integral operator of the form
\begin{equation*}
(\mathcal{L}u)(x):=\int\limits_{\Om} L(x,y)u(y)\di y,
\end{equation*}
where the kernel $L$ is an element of $L_2(\Om\times\Om)$. We
also assume that the function $L(\cdot,y)$ is compactly
supported and the relation $L(x,y)=\overline{L(y,x)}$ holds
true. Such operator satisfies the conditions (\ref{1.1}),
(\ref{1.2}). It is also $\D_{\mathbb{R}^n}$-compact and
therefore the operator $\mathcal{H}$ is self-adjoint.

\textbf{5. $\d$-potential.} The results of the general scheme
developed in the present article can be applied to the
perturbing operators not even satisfying the conditions we
impose on $\mathcal{L}$. It is possible if such operators can be
reduced by some transformations to an operator $\mathcal{L}$
satisfying needed restrictions. One of such examples is
$\d$-potential supported by a manifold. Namely, let $\G$ be a
bounded closed $C^3$-manifold in $\mathbb{R}^n$ of codimension
one and oriented by a normal vector-field
$\boldsymbol{\nu}=\boldsymbol{\nu}(\xi)$, where
$\xi=(\xi_1,\ldots,\xi_n)$ are local coordinates on $\G$. Let
$\vr$ be the distance from a point to $\G$ measured in the
direction of $\boldsymbol{\nu}$. We suppose that $\G$ is so that
the coordinates $(\vr,\xi)$ are well-defined in a some
neighbourhood of $\G$, and in this neighbourhood the mapping
$(\vr,\xi)\mapsto x$ is $C^3$-diffeomorphism. We introduce the
operator $\mathcal{H}_\G:=-\D_{\mathbb{R}^n}+b\d(x-\G)$ as
\begin{equation*}
\mathcal{H}_\G v=-\D v,\quad x\not\in\G,
\end{equation*}
on the functions
$v\in\H^2(\mathbb{R}^n\setminus\G)\cap\H^1(\mathbb{R}^n)$
satisfying the condition
\begin{equation*}
\frac{\p v}{\p \vr}\bigg|_{\vr=+0}- \frac{\p v}{\p
\vr}\bigg|_{\vr=-0}= bv\big|_{\vr=0},
\end{equation*}
where $\vr$ is $b=b(\xi)\in C^3(\G)$. We reproduce now word for
word the arguments of Example~5 in \cite[Sec. 7]{B1} to
establish
\begin{lemma}\label{lm7.1}
There exists $C^1$-diffeomorphism
$\mathcal{P}:\mathbb{R}^n\to\mathbb{R}^n$,
$\mathcal{P}=(\mathcal{P}_1,\ldots,\mathcal{P}_n)$, such that
\begin{enumerate}
\def\theenumi{\arabic{enumi}}
\item\label{lm7.1it1} The second derivatives of
$\mathcal{P}$ and $\mathcal{P}^{-1}$ exist and are piecewise
continuous.

\item\label{lm7.1it2} The function $\mathrm{p}:=\det \mathrm{P}$
and the matrix
\begin{equation*}
\mathrm{P}:=
\begin{pmatrix}
\frac{\p \mathcal{P}_1}{\p x_1} & \ldots & \frac{\p
\mathcal{P}_1}{\p x_n}
\\
\vdots & & \vdots
\\
\frac{\p \mathcal{P}_n}{\p x_1} & \ldots & \frac{\p
\mathcal{P}_n}{\p x_n}
\end{pmatrix}
\end{equation*}
satisfy the identities
\begin{equation}\label{7.4}
\begin{gathered}
\mathrm{p}^{1/2}\big|_{\vr=+0}-
\mathrm{p}^{1/2}\big|_{\vr=-0}=0,\quad
\frac{\p}{\p\vr}\mathrm{p}^{1/2}\big|_{\vr=+0}- \frac{\p}{\p\vr}
\mathrm{p}^{1/2}\big|_{\vr=-0}=b,
\\
\mathrm{P}\equiv \mathrm{E},\quad
\mathrm{p}\equiv1\quad\text{as}\quad |\vr|\geqslant\e,
\end{gathered}
\end{equation}
where $\e$ is a some small fixed number.

\item\label{lm7.1it3} The mapping $(\mathcal{U}v)(x):=
\mathrm{p}^{-1/2}v(\mathcal{P}^{-1}(x))$ is a linear unitary
operator in $L_2(\mathbb{R}^n)$ which maps the domain of the
operator $\mathcal{H}_\G$ onto $\H^2(\mathbb{R}^n)$. The
identity
\begin{equation}\label{7.5}
\mathcal{H}_{\mathcal{L}}:=\mathcal{U}\mathcal{H}_\G
\mathcal{U}^{-1}=-\D_{R^n}+\mathcal{L}
\end{equation}
holds true, where the operator $\mathcal{L}$ is  given by
(\ref{7.1}) and the supports of $b_{i,j}$, $b_i$ lie inside
$\{x: \rho\leqslant\e\}$.
\end{enumerate}
\end{lemma}

The item~\ref{lm7.1it3} of this lemma implies that the original
$\d$-potential can be reduced to a differential operator
(\ref{7.1}) without changing the spectrum. Thus, after such
transformation we can apply the results of this paper to such
perturbation as well.

The operator $\mathcal{L}$ in (\ref{7.5}) depends on the
auxiliary transformation $\mathcal{P}$. We are going to show
that the leading terms of the asymptotics expansions established
in Theorems~\ref{th1.5},~\ref{th1.7} and Corollary~\ref{cr1.6}
do not depend of $\mathcal{P}$.

We begin with Theorem~\ref{th1.5}. Let
$\mathcal{L}_k=\mathcal{L}$ for some $k$, where $\mathcal{L}$ is
from (\ref{7.5}), and $\widetilde{\psi}$ be an eigenfunction of
the operator $\mathcal{H}_{\mathcal{L}}$ associated with $\l_*$.
The corresponding elements of the matrix $\mathrm{A}_0$
introduced in Theorem~\ref{th1.5} are
\begin{equation*}
A_{i,j}^{(0)}=(\mathcal{L}u,\widetilde{\psi})_{L_2(\Om_{2\e})},
\end{equation*}
where $u=S(X_{k,r})\psi_{r,s}$, and $\Om_{2\e}:=\{x: \vr<2\e\}$.
The function $u$ satisfies the equation
\begin{equation}\label{7.7}
(\D+\l_*)u=0,\quad x\in\Om_{2\e}.
\end{equation}
The function
$\psi:=\mathcal{U}^{-1}\widetilde{\psi}=p^{1/2}\widetilde{\psi}
(\mathcal{P}(\cdot))$ is an eigenfunction of $\mathcal{H}_\G$
associated with $\l_*$ and therefore it is independent on
$\mathcal{P}$. The identities (\ref{7.4}) imply that
$\widetilde{\psi}\equiv\psi$ as $\e<\vr\leqslant 2\e$. Employing
this fact, (\ref{1.1}), (\ref{7.7}) and integrating by parts, we
obtain
\begin{equation*}
(\mathcal{L}u,\widetilde{\psi})_{L_2(\Om_{2\e})}=
(u,\mathcal{L}\widetilde{\psi})_{L_2(\Om_{2\e})}=
(u,(\D+\l_*)\widetilde{\psi})_{L_2(\Om_{2\e})}=
\int\limits_{\p\Om_{2\e}}\left(u\frac{\p\overline{\psi}}
{\p\boldsymbol{\nu}_\e}- \overline{\psi}\frac{\p
u}{\p\boldsymbol{\nu}_\e}\right)\di s,
\end{equation*}
where $\boldsymbol{\nu}_\e$ is the outward normal to
$\p\Om_{2\e}$. The last integral is independent on $\e$ since
for any $\widetilde{\e}\in(0,\e)$
\begin{gather*}
(\D+\l_*)\psi=0,\quad
x\in\Om_{2\e}\setminus\Om_{2\widetilde{\e}},
\\
0=(u,(\D+\l_*)\widetilde{\psi})_{L_2(\Om_{2\e})\setminus
\Om_{2\widetilde{\e}}}=
\int\limits_{\p\Om_{2\e}}\left(u\frac{\p\overline{\psi}}
{\p\boldsymbol{\nu}_\e}- \overline{\psi}\frac{\p
u}{\p\boldsymbol{\nu}_\e}\right)\di s-
\int\limits_{\p\Om_{2\widetilde{\e}}}
\left(u\frac{\p\overline{\psi}}
{\p\boldsymbol{\nu}_{\widetilde{\e}}}- \overline{\psi}\frac{\p
u}{\p\boldsymbol{\nu}_{\widetilde{\e}}}\right)\di s.
\end{gather*}
Using now the boundary conditions for $\psi$ on $\G$, we pass to
the limit $\e\to+0$ and obtain
\begin{equation*}
\lim\limits_{\e\to0}
\int\limits_{\p\Om_{2\e}}\left(u\frac{\p\overline{\psi}}
{\p\boldsymbol{\nu}_\e}- \overline{\psi}\frac{\p
u}{\p\boldsymbol{\nu}_\e}\right)\di
s=\int\limits_{\G}u\left(\frac{\p\overline{\psi}}
{\p\vr}\bigg|_{\vr=+0}-\frac{\p\overline{\psi}}
{\p\vr}\bigg|_{\vr=-0}\right)\di s=(u,b\psi)_{L_2(\G)}.
\end{equation*}
Thus, if an operator $\mathcal{L}_k$ describes the
$\d$-potential, the corresponding elements of the matrix
$\mathrm{A}_0$ in Theorem~\ref{th1.5} are
\begin{equation*}
A_{i,j}^{(0)}:=\big(\mathcal{S}(X_{k,r})\psi_{r,s},
\psi_{k,q}\big)_{L_2(\G)},
\end{equation*}
where $\psi_{k,q}$ are the eigenfunctions of the operator
$\mathcal{H}_\G$. In particular, if in Corollary~\ref{cr1.6} the
operator $\mathcal{H}_1$ is $\mathcal{H}_\G$, the asymptotics
expansions for $\l_i$ become
\begin{align*}
&\l_1=\l_*-\left|\big(\mathcal{L}_1
\mathcal{S}(X_{1,2})\psi_2,\psi_1\big)_{L_2(\G)}\right|
+\Odr\left(l_X^{-n+2} \E^{-2l_X\sqrt{-\l_*}}\right),\quad
l_X\to+\infty,
\\
&\l_2=\l_*+\left|\big(\mathcal{L}_1
\mathcal{S}(X_{1,2})\psi_2,\psi_1\big)_{L_2(\G)}\right|
+\Odr\left(l_X^{-n+2} \E^{-2l_X\sqrt{-\l_*}}\right),\quad
l_X\to+\infty.
\end{align*}
If under the hypothesis of Theorem~\ref{th1.7} the operator
$\mathcal{L}_1$ describes $\d$-potential, the arguments same as
given above show that the asymptotics for $\l(X)$ reads as
follows
\begin{equation*}
\l(X)=\l_*-\sum\limits_{j=2}^{m}\big(\mathcal{S}(X_{1,j})
(\mathcal{H}_j-\l_*)^{-1}\mathcal{L}_j
\mathcal{S}(X_{j,1})\psi_1,\psi_1\big)_{L_2(\G)}+
\Odr\big(l_X^{-\frac{3n-5}{2}}\E^{-3l_X\sqrt{-\l_*}}\big),
\end{equation*}
where $\psi_1$ is the eigenfunction of $\mathcal{H}_\G$. The
asymptotics for the associated eigenfunction remains the same if
by $\psi_1$ we mean the eigenfunction of $\mathcal{H}_\G$.

Suppose now that under the hypothesis of Theorem~\ref{th1.7} one
of the operators $\mathcal{L}_j$, $j\geqslant 2$, describes the
$\d$-potential. We denote
$u:=(\mathcal{H}_j-\l_*)^{-1}\mathcal{L}_j
\mathcal{S}(X_{j,1})\psi_1$. Proceeding in the same way as in
(\ref{6.1a}), we obtain
\begin{equation}\label{7.8}
\begin{aligned}
\big(\mathcal{L}_1\mathcal{S}&(X_{1,j})u,
\psi_1\big)_{L_2(\Om_1)}=\big((\D+\l_*)u,\mathcal{S}(X_{j,1})
\psi_1\big)_{L_2(\Om_{2\e})}
\\
&=\int\limits_{\p\Om_{2\e}}\left(
\mathcal{S}(X_{j,1})\overline{\psi}_1\frac{\p
u}{\p\boldsymbol{\nu}_\e}-u\frac{\p}
{\p\boldsymbol{\nu}_\e}\big(
\mathcal{S}(X_{j,1})\overline{\psi}_1\big)\right)\di s.
\end{aligned}
\end{equation}
Since $(\D+\l_*)\mathcal{S}(X_{j,1})\psi_1=0$ in $\Om_{2\e}$, it
follows that
\begin{align*}
\mathcal{L}_j\mathcal{S}(X_{j,1})\psi_1&=
(\mathcal{H}_j-\l_*)\mathcal{S}(X_{j,1})\psi_1+
(\D+\l_*)\mathcal{S}(X_{j,1})\psi_1
\\
&=\mathcal{U} \left((\mathcal{H}_\G-\l_*)\mathcal{U}^{-1}
+(\D+\l_*)\right) \mathcal{S}(X_{j,1})\psi_1.
\end{align*}
Using this relation, (\ref{7.4}), and the identity
$(\mathcal{H}_j-\l_*)^{-1}=
\mathcal{U}(H_\G-\l_*)^{-1}\mathcal{U}^{-1}$, we obtain
$u=\mathcal{U}U$, where
\begin{align*}
&U=(\mathcal{H}_\G-\l_*)^{-1}\left((\mathcal{H}_\G-\l_*)
\mathcal{U}^{-1} +(\D+\l_*)\right)\mathcal{S}(X_{j,1})\psi_1=
\widetilde{U}+U_j,
\\
&\widetilde{U}=\mathrm{p}^{1/2}\psi_1(\mathcal{P}(\cdot+X_{j,1}))
-\mathcal{S}(X_{j,1})\psi_1,
\end{align*}
and $U_j\in\H^2(\mathbb{R}^n\setminus\G)\cap\H^1(\mathbb{R}^n)$
is the unique solution to the problem
\begin{gather*}
(\D+\l_*)U_j=0,\quad x\in\mathbb{R}^n\setminus\G,
\\
\frac{\p U_j}{\p \vr}\bigg|_{\vr=+0}- \frac{\p U_j}{\p
\vr}\bigg|_{\vr=-0}= b
U_j\big|_{\vr=0}-b\mathcal{S}(X_{j,1})\psi_1\big|_{\vr=0}.
\end{gather*}
It follows from (\ref{7.4}) that $\widetilde{U}=0$, $u=U_j$ as
$\e\leqslant\vr\leqslant 2\e$. Bearing these relations in mind,
we substitute the obtained representation for $u$ into
(\ref{7.8}) and continue our calculations:
\begin{equation*}
\big(\mathcal{L}_1\mathcal{S}(X_{1,j})u,
\psi_1\big)_{L_2(\Om_1)}= \int\limits_{\p\Om_{2\e}}\left(
\mathcal{S}(X_{j,1})\overline{\psi}_1\frac{\p
U_j}{\p\boldsymbol{\nu}_\e}-U_j\frac{\p}
{\p\boldsymbol{\nu}_\e}\big(
\mathcal{S}(X_{j,1})\overline{\psi}_1\big)\right)\di s.
\end{equation*}
The right hand side of this identity is independent on small
$\e$ that allows us to pass to the limit $\e\to+0$ and obtain
\begin{align*}
\big(\mathcal{L}_1\mathcal{S}(X_{1,j})u,
\psi_1\big)_{L_2(\Om_1)}&=
\int\limits_{\G}\mathcal{S}(X_{j,1})\overline{\psi}_1\left(
\frac{\p U_j}{\p\vr}\bigg|_{\vr=+0}-\frac{\p
U_j}{\p\vr}\bigg|_{\vr=-0}\right)\di s
\\
&=\big(bU_j-b\mathcal{S}(X_{j,1})\psi_1,
\mathcal{S}(X_{j,1})\psi_1\big)_{L_2(\G)}.
\end{align*}
Finally, it leads us to the following formula
\begin{align*}
\l(X)=&\l_*- \big(bU_j-b\mathcal{S}(X_{j,1})\psi_1,
\psi_1\big)_{L_2(\G)}
\\
&- \sum\limits_{\genfrac{}{}{0pt}{}{k=2}{k\not=j}}^{m}
\big(\mathcal{L}_1\mathcal{S}(X_{1,k})
(\mathcal{H}_k-\l*)^{-1}\mathcal{L}_k
\mathcal{S}(X_{k,1})\psi_1,\psi_1\big)_{L_2(\Om_1)}+
\Odr\big(l_X^{-\frac{3n-5}{2}}\E^{-3l_X\sqrt{-\l_*}}\big),
\end{align*}
being valid as $l_X\to+\infty$ if the operator $\mathcal{H}_j$
describes the $\d$-potential.


\section*{Acknowledgments}

The research was supported by \emph{Marie Curie International
Fellowship} within 6th European Community Framework
(MIF1-CT-2005-006254). The author is also supported by the
Russian Foundation for Basic Researches (Nos. 06-01-00138,
05-01-97912-r\_agidel) and by the Czech Academy of Sciences and
Ministry of Education, Youth and Sports (LC06002).


%

\end{document}